\documentclass[preprint2]{aastex}
\slugcomment{{\sc Accepted to ApJ:} Sept. 11, 2015}

\usepackage{hyperref}
\usepackage{natbib}
\usepackage{amsmath}

\newcommand{\myemail}{ceharmanjr@psu.edu}

\newcommand{\2}{$_{2}$}

\newcommand{\EXP}[2]{e$^{\frac{#1}{#2}}$}
\newcommand{\tbdyt}[2]{ $\Bigg \{$
\parbox{5cm}{
k$_{0}$=#1 \\ 
k$_{\infty}$=#2  
}
}
\newcommand{\tbdy}[4]{ $\Bigg \{$
\parbox{5cm}{
k$_{0}$=#1 $\cdot$(T/300)$^{#2}$ \\ 
k$_{\infty}$=#3 $\cdot$(T/300)$^{#4}$ 
}
}
\newcommand{\tbdyg}[3]{ $\Bigg \{$
\parbox{6cm}{
FC=#1 \\ 
k$_{0}$=#2  \\
k$_{\infty}$=#3
}
}

\newcommand{\refeq}[1]{eq. \ref{#1}}
\newcommand{\reffig}[1]{Fig. \ref{#1}}
\newcommand{\reftable}[1]{Table \ref{#1}}

\newcommand{\e}[1]{$\times10^{#1}$}

\shorttitle{Abiotic O\2 Levels}
\shortauthors{Harman, Schwieterman, Schottelkotte, and Kasting}

\begin{document}

\title{Abiotic O\2 Levels on Planets around F, G, K, and M Stars:\\ Possible False Positives for Life?}

\author{C. E. Harman\altaffilmark{1,2,3}}
\affil{Geosciences Department, Pennsylvania State University, University Park, PA 16802}
\email{\myemail}

\author{E. W. Schwieterman\altaffilmark{2}}
\affil{Astronomy Department, University of Washington, Seattle, Washington 98195}

\author{J. C. Schottelkotte}
\affil{Astronomy Department, Pennsylvania State University, University Park, PA 16802}

\and

\author{J. F. Kasting\altaffilmark{1,2,3}}
\affil{Geosciences Department, Pennsylvania State University, University Park, PA 16802}

\altaffiltext{1}{Pennsylvania State Astrobiology Research Center}
\altaffiltext{2}{NASA Astrobiology Institute -- Virtual Planetary Laboratory}
\altaffiltext{3}{Center for Exoplanets and Habitable Worlds}

\begin{abstract}
In the search for life on Earth-like planets around other stars, the first (and likely only) information will come from the spectroscopic characterization of the planet's atmosphere. Of the countless number of chemical species terrestrial life produces, only a few have the distinct spectral features and the necessary atmospheric abundance to be detectable. The easiest of these species to observe in Earth's atmosphere is O\2 (and its photochemical byproduct, O$_{3}$). But O\2 can also be produced abiotically by photolysis of CO\2, followed by recombination of O atoms with each other. CO is produced in stoichiometric proportions. Whether O\2 and CO can accumulate to appreciable concentrations depends on the ratio of far-UV to near-UV radiation coming from the planet's parent star and on what happens to these gases when they dissolve in a planet's oceans. Using a one-dimensional photochemical model, we demonstrate that O\2 derived from CO\2 photolysis should not accumulate to measurable concentrations on planets around F- and G-type stars. K-star, and especially M-star planets, however, may build up O\2 because of the low near-UV flux from their parent stars, in agreement with some previous studies. On such planets, a `false positive' for life is possible if recombination of dissolved CO and O\2 in the oceans is slow and if other O\2 sinks (e.g., reduced volcanic gases or dissolved ferrous iron) are small. O$_{3}$, on the other hand, could be detectable at UV wavelengths ($\lambda<$ 300 nm) for a much broader range of boundary conditions and stellar types.
\end{abstract}


\section{Introduction}
\label{intro}

The discovery of Earth-sized planets orbiting within the habitable zones of various stars using ground-based radial velocity measurements \citep{anglada2012} and transit observations from NASA's Kepler Space Telescope \citep{borucki2012,borucki2013,quintana2014} has led to increased interest in the question of whether other Earth-like planets exist and whether any of them might be inhabited. If Earth- and super-Earth-sized planets can be found around nearby stars with the Transiting Exoplanet Survey Satellite (TESS) \citep{ricker2014}, we may be able to analyze their atmospheres with transit spectroscopy using the upcoming James Webb Space Telescope (JWST) \citep{deming2009}. The next generation of space telescope mission concepts, such as the Advanced Technology Large Aperture Space Telescope (ATLAST) or the Large Ultraviolet/Optical/Infrared Telescope (LUVOIR) \citep[e.g.,][]{postman2010,france2015}, could use reflection spectroscopy to characterize many more such planets. Future large (30-40 m aperture) ground-based telescopes could also characterize some of the nearest detected terrestrial planets \citep{kawahara2012,snellen2013}. More focused mission concepts, such as NASA's \href{http://exep.jpl.nasa.gov/archive/}{Terrestrial Planet Finder} (TPF-C) or Webster Cash's \href{http://newworlds.colorado.edu/}{New Worlds Observer Mission} (also known as TPF-O), might also characterize a large number of Earth-like planets. Earth-like exoplanets could be studied at longer wavelengths using emission spectroscopy by interferometers such as NASA's TPF-I or ESA's Darwin mission \citep{cockell2009}. None of these missions is being actively pursued by NASA at the moment but they could be selected for development several years from now if they receive a favorable rating in the next Astronomy and Astrophysics Decadal Survey. Smaller probe-class concepts featuring a coronagraph (\href{http://exep.jpl.nasa.gov/stdt/Exo-C_Final_Report_for_Unlimited_Release_150323.pdf}{Exo-C}) or a starshade (\href{http://exep.jpl.nasa.gov/stdt/Exo-S_Starshade_Probe_Class_Final_Report_150312_URS250118.pdf}{Exo-S}) would be a low-cost stepping stone from JWST to a dedicated TPF mission, and could fly in concert with the next UVOIR space telescope.

\subsection{Disequilibrium pairs of gases as a biosignature}
\label{diseq}

Because atmospheric characterization of exoplanets can be done simultaneously with direct detection, the question immediately arises as to what gases, or combination of gases, might provide evidence for the existence of life on such a planet. A capable review of this subject has been given by \citet{seager2015}. We provide our own brief synopsis here. The literature on remote life detection extends back well before the exoplanet era. \citet{lovelock1965} first suggested that the presence of oxygen in a planet's atmosphere, alongside hydrocarbons, would constitute a reliable biosignature. Later, \citet{lippincott1967} extended this to specific gas pairs such as O\2 and CH$_{4}$ or N\2O. These gases are many orders of magnitude out of thermodynamic equilibrium with each other, and all have sources that are dominantly biological. \citet{lederberg1965} had already proposed that extreme thermodynamic disequilibrium, in general, would be good evidence for life. This concept has been explored recently by \citet{bains2012} (see also \citet{seager2012,seager2013a,seager2013b} and \citet{krissansen2015}), but no additional, potentially observable, disequilibrium redox pair has been described. Moreover, photochemical modelers \citep[e.g.,][]{kasting1990,kasting2014,zahnle2008} have shown that CO-rich atmospheres can arise under a variety of circumstances. CO is a high free-energy compound that would be well out of thermodynamic equilibrium in almost any plausible abiotic terrestrial atmosphere. But, for this same reason, it should be an excellent source of metabolic energy for microbes \citep{kharecha2005,kasting2014}, and so its presence in a planet's atmosphere might actually be considered an anti-biosignature \citep{zahnle2008}. We thus agree with \citet{seager2015} that the use of thermodynamic disequilibrium as a biosignature cannot be easily generalized; however, the more specific life detection criterion proposed by \citeauthor{lovelock1965} and \citeauthor{lippincott1967} is still useful.

Herein lies a difficulty, however. O\2 is highly abundant (21\% by volume) in Earth's atmosphere and should be readily detectable by a first-generation TPF-type mission by way of its `A' absorption band at 760 nm \citep{owen1980,desmarais2002}. But CH$_{4}$ is present at only 1.7 ppmv and would likely not be detectable by a first-generation TPF-type telescope, despite the fact that it absorbs in both the visible and near-IR. Other, clearly biogenic, gases such as methyl chloride (CH$_{3}$Cl) and dimethyl sulfide (DMS) are also unlikely to be detectable by TPF \citep{domagal2011,seager2015}. This leads to the question that is the main topic of this paper: Can O\2 by itself be considered a biosignature and, if so, under what conditions is this interpretation valid? 

\subsection{O\2 by itself as a biosignature}
\label{o2bio}

\subsubsection{Historical development}
\label{history}

We begin by acknowledging that many previous researchers have suggested, directly or indirectly, that O\2 is \emph{not} a good bioindicator. In a series of papers written during the 1960's, \citet{berkner1964,berkner1965,berkner1966,berkner1967} calculated that atmospheric O\2 could build up to as high as 10$^{-3}$ PAL (times the Present Atmospheric Level) as a consequence of photolysis of H\2O followed by escape of hydrogen to space. This amount of O\2 is just on the verge of being detectable by TPF, primarily through its photochemical byproduct, O$_{3}$ \citep{leger1993,segura2003}. And, as we discuss further below, this is also the O\2 level at which confusion with biologically produced O\2 begins to become an issue. Several years later, \citet{brinkmann1969} predicted prebiotic O\2 levels as high as 0.27 PAL, which would certainly be detectable by currently envisioned technology. Both studies, however, were performed before the process of hydrogen escape from Earth's atmosphere was fully understood. We know now that the rate of hydrogen escape is limited by diffusion through the homopause, near 100 km \citep{hunten1973,walker1977}, and this diffusion rate, in turn, depends on the total hydrogen mixing in the stratosphere: f$_{tot}$(H) = f(H) + 2 f(H\2) + 2 f(H\2O) + 4 f(CH$_{4}$) + \ldots \citep[see, e.g.,][]{kasting2003}. The modern Earth's stratosphere is very dry (f(H\2O) $\approx$ 3-5 ppmv), and reduced gases are also at or below the ppmv level, so the current rate of O\2 production from hydrogen escape is relatively slow.

More recent claims of high abiotic O\2 levels have been made by researchers using improved photochemical models and more realistic assumptions about hydrogen escape. \citet{canuto1982,canuto1983} predicted high abiotic O\2 levels from CO\2 photolysis by a highly UV-active young Sun. Their assumed UV levels were unrealistically high for the early Earth, and their model did not obey redox balance (more on this below), so their apparent false positive was probably not correct \citep{kasting1984}. \citet{rosenqvist1995} predicted high abiotic O\2 concentrations, but their model also failed to consider redox balance \citep{kasting1995}. \citet{selsis2002} predicted high abiotic O\2 levels in dense CO\2 atmospheres, but their model suffered from this same problem \citep{segura2007}. Given the critical importance of redox balance to abiotic O\2 levels and the absence of any consideration of it in numerous papers in the literature, we include a discussion of this topic below.

\subsubsection{Planets that suffer extensive water loss}
\label{waterloss}

One class of apparent false positives for life involves planets that lose a lot of water. A good example is a planet interior to the habitable zone that loses its water by way of a runaway or moist greenhouse \citep{kasting1988,kasting1997,kasting2010,wordsworth2013}. A fully vaporized Earth ocean would have a surface pressure of $\sim$270 bar. Thus, if such a steam atmosphere was photolyzed by stellar UV, and if the hydrogen escaped to space, the planet could be left with a 240-bar O\2 atmosphere, having lost just over 10 percent of the ocean mass. But this type of false positive is not too worrisome because we would not see H\2O absorption bands, unless we were unlucky enough to observe the planet during the brief period in which the ocean had evaporated but not yet been completely lost; thus, we would likely deduce that the observed O\2 was the product of a runaway greenhouse. Exogenous water delivered later in the planet's lifetime should be also be short-lived, as the planet should be even more susceptible to water loss when its parent star is older and brighter. A similar process could happen to a planet in the middle of the HZ if the background pressure of noncondensable gases such as N\2, CO\2, and Ar was small, because H\2O would then be a major atmospheric constituent \citep{wordsworth2014}. Water loss would continue until the background gas (in this case, O\2) built up to $\sim$0.2 bar. This type of false positive would be more difficult to distinguish spectroscopically, but might be eventually identified by looking for O\2-O\2 and N\2-N\2 features that could constrain the atmospheric partial pressures of these two gases \citep{misra2014,schwieterman2015}.

A recent variant on the runaway greenhouse concerns planets orbiting M stars. \citet{luger2015b} (see also \citet{ramirez2014} and \citet{tian2015}) have shown that because of the long, highly luminous, pre-main-sequence lifetime of M stars, planets orbiting within their habitable zones could lose most or all of their water early in their histories, with accompanying buildup of atmospheric O\2. This would be true for terrestrial planets around M stars with long pre-main sequence lifetimes, regardless of their (largely unconstrained) initial volatile abundances \citep[e.g.,][]{raymond2004,raymond2007a,raymond2007,raymond2009,ciesla2015}. Such planets should also lack strong H\2O absorption features, much as Venus does today, and so the O\2 signal could probably be discounted as evidence for life (Venus' 30 ppmv of H\2O is detectable spectroscopically, but only by looking at high spectral resolution through near-IR `windows' on the planet's nightside \citep{meadows1996} -- Venus' H\2O is not detectable at TPF-type spectral resolution \citep{meadows2005}). \citet{hamano2013} have suggested that planets that start off with a runaway greenhouse atmosphere could have their O\2 absorbed by the magma ocean, perhaps removing the possibility of a false positive altogether. But one can imagine M-star planets that are resupplied with water at some later time by some process akin to the Late Heavy Bombardment in our own Solar System, or that migrate into the HZ after their host star evolves onto the main sequence. Hence, we should still be concerned about false positive on M-star planets, as we discuss further below.

\subsubsection{Planets with frozen surfaces}
\label{frozen}

A slightly less obvious false positive is a planet just beyond the outer edge of the habitable zone, like early Mars, but with roughly twice Mars' mass \citep{kasting1997,kasting2010}. The frozen surface would not take up much oxygen once the surface layer itself was oxidized, and the planet should be large enough to hold onto its atmosphere, but perhaps not large enough to maintain volcanic outgassing of reduced species like H\2. In that case, photodissociation of H\2O in the planet's atmosphere, followed by escape of hydrogen to space, could conceivably cause substantial amounts of O\2 to accumulate, just as on the Venus-like planet described above. The martian atmosphere itself contains $\sim$0.1\% O\2, produced by this process. Indeed, Mars' O\2 concentration would probably be even higher except that the planet's low gravity allows oxygen to escape by way of nonthermal processes such as dissociative recombination of ions, e.g., O\2$^{+}$ + e $\rightarrow$ O + O, in which the resultant O atoms have sufficient kinetic energy to escape. Most of these nonthermal loss processes for oxygen would cease to operate on a planet twice as massive as Mars, allowing O\2 to accumulate almost indefinitely. The planet's upper atmosphere would be dry, making it unlikely to be distinguishable using transit transmission spectroscopy, but this type of false positive could probably be identified with direct imaging because the planet surface would be cold, and hence its lower atmosphere should be deficient in H\2O. 

\subsubsection{More troubling false positives: low-outgassing planets and planets around M stars}
\label{falsea}

\begin{figure}
\includegraphics[width=\textwidth]{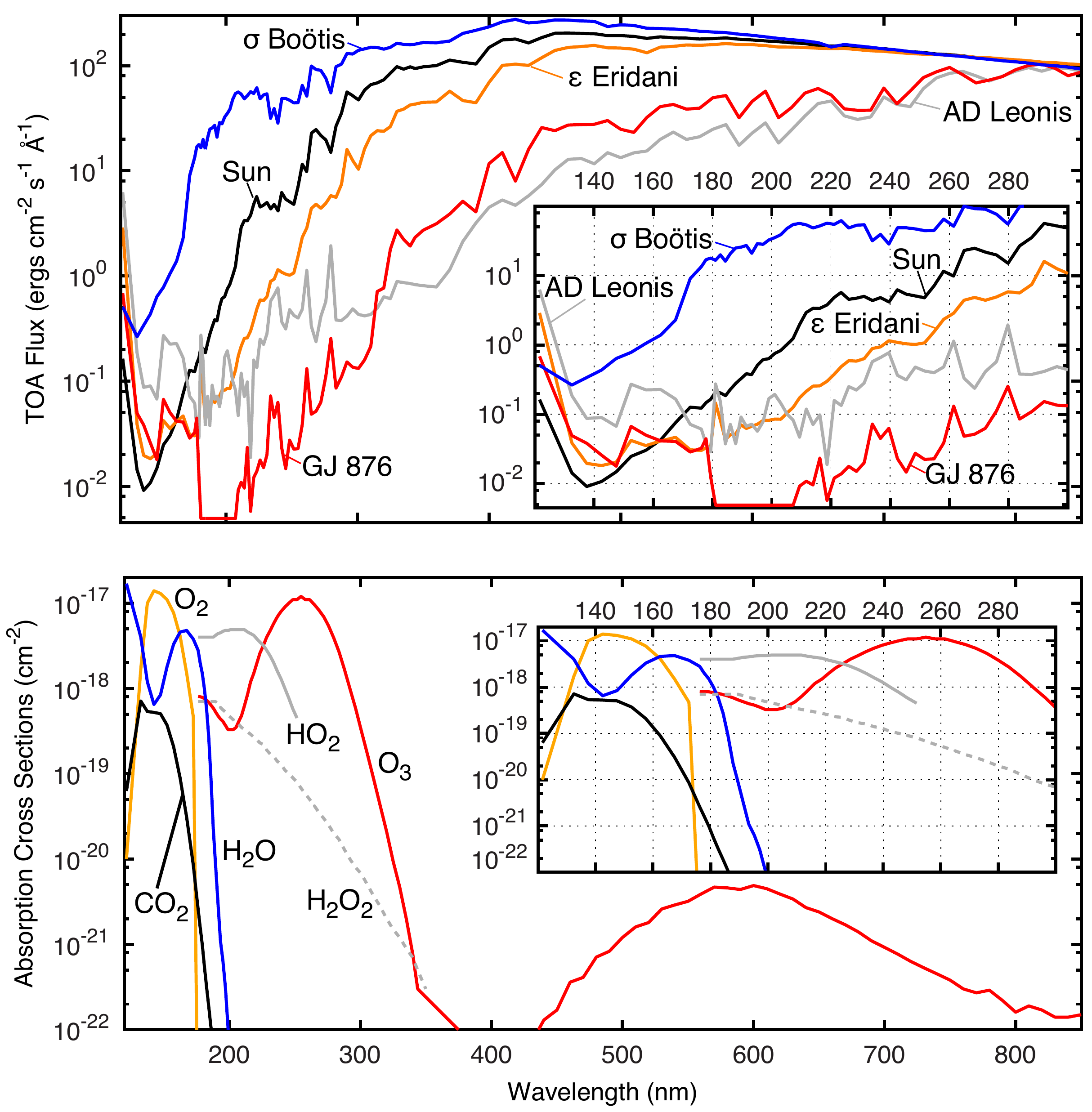}
\caption{The top-of-atmosphere stellar flux for a planet located at 1 AU equivalent distance (receiving 1,360 W m$^{-2}$ of stellar radiation). The planets discussed here are located at 1.3 AU equivalent, which corresponds to a $\sim$40\% reduction in flux. \label{specx}}
\end{figure}

\begin{figure}
\includegraphics[width=\textwidth]{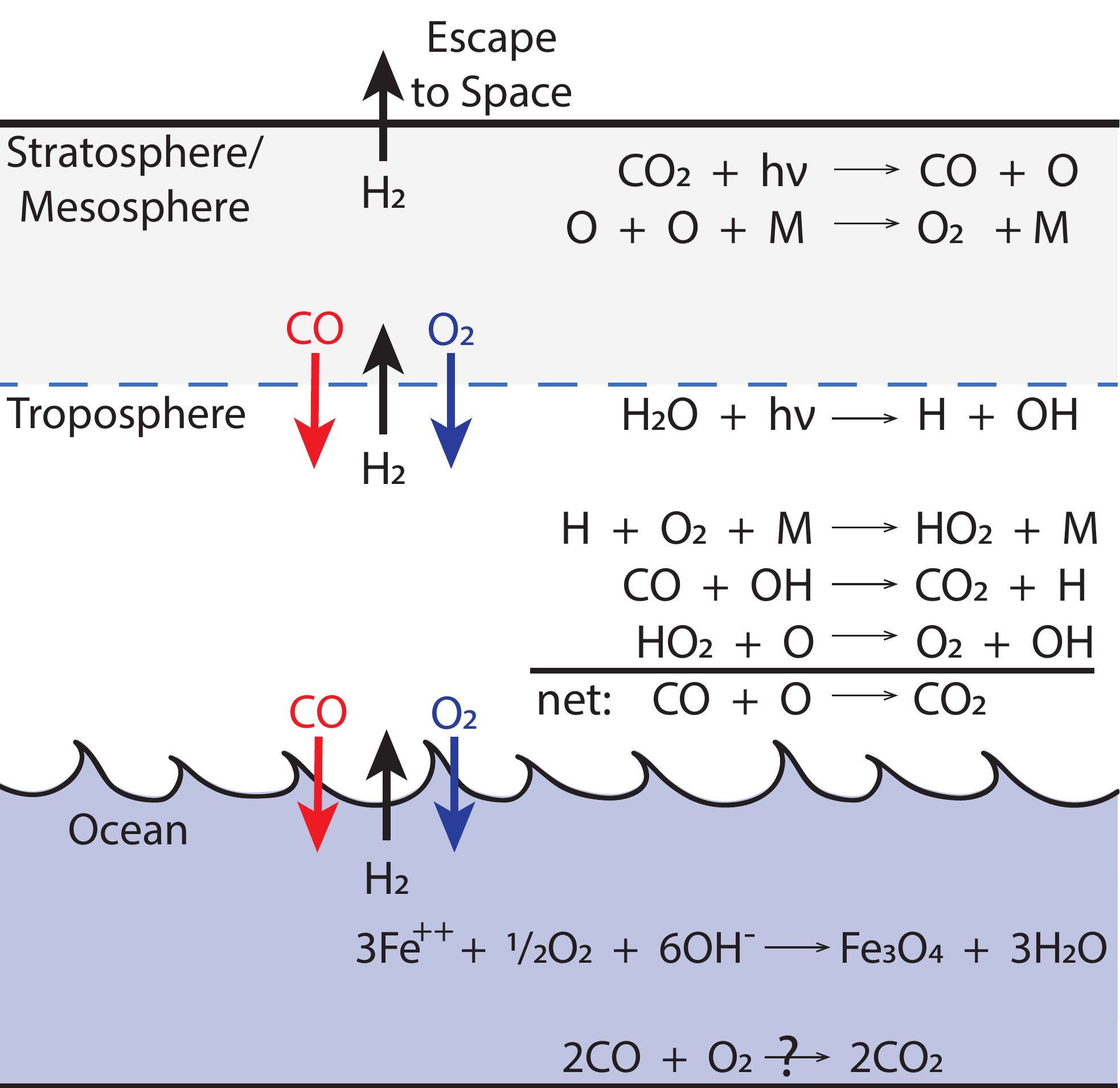}
\caption{Summarizing the reactions that control abiotic O\2 levels in terrestrial atmospheres and oceans. \label{scheme}}
\end{figure}

Recently, several researchers have suggested the existence of more troubling false positives that would be harder to identify. It is these studies that motivate the present paper. \citet{hu2012} concluded that CO\2-dominated worlds around Sun-like stars may build up oxygen if the outgassing rates of reduced gases (such as methane or hydrogen) are small. Their model included an atmospheric hydrogen budget, along with realistic assumptions about the rate of hydrogen escape. More recently, \citet{tian2014} have suggested that CO\2-rich planets orbiting M stars may build up significant amounts of O\2 because of the higher ratio of far ultraviolet to near ultraviolet (FUV/NUV) stellar radiation. Far-UV radiation ($\lambda<$200 nm) can photolyze CO\2, producing CO and O atoms:
\begin{center}
\begin{tabular}{ccc}
 CO\2  +  h$\nu$ & $\rightarrow$ & CO  +  O
\end{tabular}
\end{center}
Direct recombination of CO with O is slow, however, because that reaction is spin-forbidden. Hence, O atoms recombine with each other to form O\2:
\begin{center}
\begin{tabular}{ccc}
 O  +  O  +  M  & $\rightarrow$ & O\2  +  M
\end{tabular}
\end{center}
Here, `M' is a third molecule, needed to carry of the excess energy from the collision. Recombination of O or O\2 with CO can be catalyzed by the byproducts of H\2O photolysis, which occurs at wavelengths $<$ 240 nm. (All of these photochemical models, including the independent one by Hu et al., extrapolate the measured H\2O cross sections from 196 nm out to 240 nm, which is the energy threshold for breaking the H\2O bond \citep{kasting1981}. The H\2O absorption cross section is low, but finite, at these wavelengths, giving rise to substantial H\2O photolysis all the way down to the planet's surface.) For example, one such catalytic cycle (originally proposed by \citet{donahue1969} and \citet{mcelroy1970}) begins with
\begin{center}
\begin{tabular}{ccc}
 H\2O  +  h$\nu$ & $\rightarrow$ & OH  +  H
\end{tabular}
\end{center}
followed by
\begin{center}
\begin{tabular}{cccc}
 &CO  +  OH & $\rightarrow$ & CO\2  +  H \\
 &O\2 + H + M & $\rightarrow$ & HO\2  +  M \\
 &HO\2 + O & $\rightarrow$ & O\2 + OH \\ \hline
 net: & CO + O & $\rightarrow$ & CO\2
\end{tabular}
\end{center}
The wavelength regions for photolysis of H\2O and CO\2 overlap with each other (\reffig{specx}); however, only H\2O can be photolyzed longward of 200 nm, and so the efficiency of such catalytic cycles depends critically on the flux of stellar near-UV radiation. We have summarized these reactions in \reffig{scheme}.


Domagal-Goldman et al. suggested that total far-UV ($<$200 nm) fluxes in excess of solar levels are responsible for the generation of abiotic O\2 and O$_{3}$. In contrast, Tian et al. suggest that the higher ratio of far-UV ($<$200 nm) to mid- and near-UV (200-400 nm, hereafter referred to simply as the near-UV) radiation coming from an M star (as compared to the Sun) is responsible for increased O\2 and O$_{3}$ production. For the spectra used in this study, the difference is centered at $\sim$170 nm, characterized by a marked change in flux for the F, G, and K stars, with little to no change in flux for the M stars (\reffig{specx}). Many M stars are highly active and so emit an excess of far-UV radiation from their chromospheres. Near-UV radiation is absorbed within the photosphere by molecules such as TiO, and so the stellar emission is sub-blackbody at these wavelengths. As we will show later, the NUV/FUV plays a critical role in determining the steady state concentrations of CO and O\2, rooted in the photochemistry outlined above.

Returning to the model comparison, we note that the Tian et al. model also included an atmospheric hydrogen budget. But when \citet{domagal2014} repeated these calculations, they got mixed results: these authors agreed with Hu et al. and disagreed with Tian et al. This outcome was somewhat surprising because the Hu et al. photochemical model was developed independently, but the Domagal-Goldman et al. and Tian et al. models were offshoots of the same Kasting group model. Thus, one might have expected the agreement/disagreement to be the other way around. Here, we compare our own (Kasting group) model with all three of these other models in an attempt to identify the reasons for the discrepancies and to determine which predictions are robust and which are model-dependent. 

\section{What Constitutes a False Positive for Life?}
\label{falseb}

Before proceeding further, we should step back and decide what exactly constitutes a false positive for life. Oftentimes, this has been defined as detectable atmospheric O\2 or O$_{3}$ concentrations that are produced abiotically. But this definition is unsatisfying because the term `detectable' depends critically on the technology used for detection. Already, three different classes of telescopes have been advertised as having the ability to characterize Earth-like planets: i) JWST, ii) the TPF-type telescopes mentioned above, and iii) large ground-based telescopes. Each of these telescopes has different exoplanet characterization capabilities, and none of them may be able to measure atmospheric O\2 concentrations well below that of present Earth. But if any of these instruments were to succeed in identifying a promising, Earth-like planet, it is almost guaranteed that some later generation of astronomers would design a more capable telescope to follow up on those observations. So, a more general way to think about this problem is that any abiotic O\2 concentration that exceeds measured, or inferred, biotically produced O\2 levels on Earth should be considered as a potential false positive. Whether it would be detectable by a first-generation TPF-type telescope is an interesting, but separate, question.

We know very accurately what Earth's O\2 concentration is today -- close to 21 percent by volume (1 PAL). But we have good reason to believe that O\2 concentrations were much lower earlier in Earth's history, including times during which oxygenic photosynthesis is believed to have been operative. Oxygenic photosynthesis was invented by cyanobacteria during the Archean Eon, possibly as early as 3.0 billion years ago (Ga) \citep{crowe2013}, although free O\2 did not begin to accumulate in the atmosphere until around 2.5 Ga (the so-called Great Oxidation Event, or GOE). Some free O\2 was present during the Archean as a consequence of photolysis of CO\2, but this O\2 was largely confined to the stratosphere and was not appreciably more abundant than on the prebiotic Earth \citep{kasting1979,kasting2003}. 

O\2 rose to appreciable concentrations in the lower atmosphere during the GOE, as evidenced by the disappearance of sulfur mass-independent fractionation in sedimentary rocks, along with other geologic O\2 indicators \citep{holland2006}. During most of the ensuing Proterozoic Eon (2.5-0.54 Ga), the atmosphere was oxygenated, but O\2 concentrations are thought to have remained significantly lower than today. Until recently, the hypothesized range of values for Proterozoic O\2 was 0.01-0.5 PAL \citep{kump2008}. These values have always been uncertain, however, and a recent study by \citet{planavsky2014} suggests that mid-Proterozoic O\2 concentrations were at most 10$^{-3}$ PAL. This estimate is uncertain, as it depends on the complex geochemical behavior of the element chromium, which has only recently been suggested as an O\2 indicator. However, this is currently our best estimate for Proterozoic O\2, and we adopt it here as a lower bound for a biotically sustained O\2 concentration.

The Planavsky et al. results, if correct, imply that O\2 would have been difficult to detect directly during much of Earth's history by a first-generation, optical, TPF-type telescope. Direct detection of the O\2 A band by such a device requires O\2 concentrations of $\sim$0.01 PAL \citep{desmarais2002,segura2003}. In the visible wavelength region, then, the Proterozoic Earth may be an example of an inhabited world with no detectable atmospheric biosignatures \citep{cockell2014}. However, a space telescope operating in the thermal-infrared could look for the ozone band at 9.6 $\mu$m, and that signal should be detectable for an O\2 level as low as 10$^{-3}$ PAL \citep{leger1993,segura2003}. Alternatively, a large UVOIR telescope could potentially observe O$_{3}$ features in the UV. So, one might be able to identify an analog to the Proterozoic Earth using a future space-based telescope operating outside of the visible wavelength region.

\section{Atmospheric and Global Redox Balance}
\label{atmobal}

As mentioned earlier, many early attempts to calculate abiotic atmospheric O\2 concentrations did not properly account for redox balance. Redox balance is simply conservation of free electrons. Or, to say it another way, when one species is oxidized (loses electrons), another species must be reduced (gain electrons). Both the atmosphere itself and the combined atmosphere-ocean system must satisfy redox balance over time scales of tens of thousands to millions of years; otherwise, their compositions would change. The relevant time scale for redox balance in a weakly reduced early Earth-type atmosphere is $\sim$30,000 yrs \citep{kasting2013}. This is readily calculated by dividing the column depth of H\2 by the diffusion-limited escape rate of hydrogen. The relevant time scale for Earth's O\2-rich modern atmosphere is about 2 million years \citep{holland2002}. This is the reservoir size of atmospheric O\2 divided by the rate of consumption of O\2 by surface weathering and oxidation of reduced volcanic gases.

Atmospheric redox balance can be explicitly calculated in a photochemical model. Indeed, if the model is written self-consistently (\emph{i.e.}, conserving mass and with stoichiometrically balanced chemical reactions), it \emph{must} conserve redox balance because free electrons can neither be gained nor lost. To track redox balance, we assign each species a redox coefficient based on the number of H\2 molecules that are generated or consumed when converting that species to a redox-neutral one. Following \citet{kasting2012} and \citet{kasting2013}, we define H\2O, SO\2, CO\2, and N\2 as neutral species for H- and O-, S-, C-, and N-bearing gases, respectively. The redox coefficient for each species can be found in \reftable{species} in the Model Description (section \ref{model}). For example, H\2S has a redox coefficient of +3, because H\2S + 2H\2O $\rightarrow$ SO\2 + 3H\2, that is, one H\2S molecule converted to SO\2 liberates 3 molecules of H\2. Another example would be H\2O\2, which has a redox coefficient of -1, as it consumes 1 molecule of H\2 when converted to water vapor (H\2O\2 + H\2 $\rightarrow$ 2H\2O).  Under these assumptions, the atmospheric redox balance can be written as
\begin{equation}
\Phi_{out}(Red) + \Phi_{rain}(Ox) = \Phi_{esc}(H_{2}) + \Phi_{rain}(Red) \label{atmo1}
\end{equation}
Here, $\Phi_{out}(Red)$ is the outgassing rate of reduced species in all chemical forms, weighted by their stoichiometric coefficient for H\2 production
\begin{align}
\Phi_{out}(Red)&= \Phi_{out}(H_{2})  + \Phi_{out}(CO)  \nonumber\\+& 4\Phi_{out}(CH_{4})  + 3\Phi_{out}(H_{2}S) + \ldots \label{atmo2}
\end{align}
$\Phi_{esc}(H_{2}) $ is the hydrogen escape rate to space, assumed to be given by the diffusion-limited flux
\begin{equation}
\Phi_{esc}(H_{2}) \approx 2.5\times10^{13}f_{T}(H_{2}) \label{atmo3}
\end{equation}
and $f_{T}(H_{2})$ ($= 0.5f(H) + f(H_{2}) + f(H_{2}O) + 2f(CH_{4}) + \ldots$) is the total volume mixing ratio of hydrogen in all its chemical forms. Meanwhile, $\Phi_{rain}(Ox)$ and $\Phi_{rain}(Red)$ are the combined rainout and surface deposition rates of oxidants and reductants, respectively. We treat rainout and surface deposition together because both processes are assumed to result in transfer of gases from the atmosphere to the ocean.

Global redox balance refers to redox balance within the combined atmosphere-ocean system. In the notation of \citet{kasting2013}, this can be expressed as
\begin{align}
\Phi_{out}(Red) &+ \Phi_{OW} + \Phi_{burial}(CaSO_{4}) +\nonumber\\ \Phi_{burial}(Fe_{3}O_{4})& =\Phi_{esc}(H_{2}) + 2\Phi_{burial}(CH_{2}O) \nonumber\\+ & 5\Phi_{burial}(FeS_{2}) \label{atmo4}
\end{align}
H\2 sources are on the left-hand side of this equation; H\2 sinks are on the right. Here, $\Phi_{OW}$ represents oxidative weathering of the continents and seafloor and $\Phi_{burial}(i)$ is the burial rate of species $i$ in sediments. Important species are gypsum (CaSO$_{4}$), magnetite (Fe$_{3}$O$_{4}$), organic matter (CH\2O), and pyrite (FeS\2). We have written them in this manner to show the stoichiometry explicitly. Other authors \citep[e.g.,][]{catling2005} have used different notation. Many authors \citep[e.g.,][]{holland2002,holland2009} compute the global redox budget in units of O\2, rather than H\2, and they adopt sulfate (SO$_{4}^{=}$) as the reference oxidation state for sulfur (although frequently these assumptions are not stated explicitly). These latter parameters are good choices for the O\2-rich modern atmosphere but are less well suited for low-O\2 atmospheres.

Unlike atmospheric redox balance, a photochemical model will not compute global redox balance automatically because it does not include an ocean. But global redox balance can be imposed by making certain assumptions about the magnitude of the terms in \refeq{atmo4} and then adjusting the lower boundary conditions of the photochemical model so that the retained terms are in balance. On an abiotic planet, oxidative weathering and all sediment burial terms are expected to be small. If we neglect these terms, then \refeq{atmo4} simplifies to
\begin{equation}
\Phi_{out}(Red) = \Phi_{esc}(H_{2})  \label{atmo5}
\end{equation}
\emph{i.e.}, volcanic outgassing of reduced gases must be balanced by escape to space. By combining this expression with \refeq{atmo1}, we see that our model must also satisfy the equation
\begin{equation}
\Phi_{rain}(Ox) =  \Phi_{rain}(Red) \label{atmo6}
\end{equation}
that is, rainout plus surface deposition of oxidants must equal rainout plus surface deposition of reductants. Another way to think of this is that, if consumption of oxidants by the crust and burial of reductants in sediments are neglected, oceanic redox balance requires that the inputs of oxidants and reductants from the atmosphere must be in balance. We will return to this thought below because it is key to understanding the modeling approach taken here, as well as those adopted by \citet{tian2014} and \citet{domagal2014}. Until those articles appeared, most papers regarding false positives (including some written by us, e.g., \citet{segura2007}), have simply overlooked the concept of global redox balance.

\section{Model Description} \label{model}

\subsection{Model parameters and atmospheric chemistry}
\label{modelparam}

To compare with other recent false positive calculations, we used a 1-D, horizontally averaged, photochemical model that shares its heritage with that of \citet{segura2007}, \citet{tian2014}, and \citet{domagal2014}. The model has 30 long-lived species (\reftable{species}), 15 short-lived species, and one aerosol species (sulfate). N\2 is treated as inert and is given a constant vertical mixing ratio set by the abundance of other gases so as to ensure a 1-bar surface pressure. This keeps us out of the parameter space in which the background pressure is low and water loss is rapid \citep{wordsworth2014}. Chemical steady state is calculated via 215 reactions, including the photolysis of key atmospheric constituents. The full reaction list is given in the Appendix.

\begin{deluxetable}{ccc}
\tablewidth{0pt}
\tablecaption{Chemical species \label{species}}
 \tablehead{
 \colhead{Chemical} & \colhead{Deposition} & \colhead{Redox} \\
 \colhead{formula} & \colhead{velocities (cm s$^{-1}$)} & \colhead{coefficient}
 }
 \startdata
O & 1. & -1 \\\hline
O\2 & (0-1.4\e{-4})& -2 \\\hline
H\2O & fixed\tablenotemark{a} & 0 \\\hline
H & 1. & 0.5 \\\hline
OH & 1. & -0.5 \\\hline
HO\2 & 1. & -1.5 \\\hline
H\2O\2 & 0.5 & -1 \\\hline
H\2 & variable\tablenotemark{b} & 1 \\\hline
CO\2 & fixed\tablenotemark{c} & 0 \\\hline
CO & (0-1.2\e{-4}) & 1 \\\hline
HCO & 1. & 1.5 \\\hline
H\2CO & 0.1 & 2 \\\hline
CH$_{4}$ & 0. & 4 \\\hline
CH$_{3}$ & 1. & 3.5 \\\hline
C\2H$_{6}$ & 1.0\e{-5} & 7 \\\hline
NO & 3.0\e{-4} & -1 \\\hline
NO\2 & 3.0\e{-3} & -2 \\\hline
HNO & 1. & -0.5 \\\hline
H\2S & 0.015 & 3 \\\hline
HS & 3.0\e{-3} & 2.5\\\hline
S & 1. & 2 \\\hline
HSO & 1. & 1.5 \\\hline
SO & 3.0\e{-4} & 1 \\\hline
SO\2 & 1. & 0 \\\hline
NH$_{3}$ & 0. & 1.5 \\\hline
NH\2 & 0. & 1 \\\hline
N & 0. & 0 \\\hline
N\2H$_{4}$ & 0. & 2 \\\hline
N\2H$_{3}$ & 0. & 1.5 \\\hline
H\2SO$_{4}$ & 0.2 & -1 \\\hline
\enddata
\tablecomments{
\tablenotemark{a}{H\2O volume mixing ratio is fixed};
\tablenotemark{b}{H\2 is used to balance the global redox budget - see the text for a description};
\tablenotemark{c}{The CO\2 mixing ratio is fixed between 0.05 and 0.9}.}
\end{deluxetable}

We did not attempt to calculate self-consistent temperature profiles for different model atmospheres. Rather, we assumed temperatures comparable to those found in the models with which we wish to compare \citep[e.g.,][]{hu2012,tian2014}. Similarly, we employed two different eddy diffusion profiles: the first from Hu et al. for our 90\% CO\2 cases, and the second from Tian et al. for our 5\% CO\2 cases. The eddy diffusion profile used by \citet{tian2014} results in a lowered homopause (located around 75 km), which accounts for the decreasing CO\2 concentrations in the upper atmosphere of their model. 

\subsection{Stellar spectra}
\label{stellar}

Some of our calculations involve Earth-like exoplanets orbiting F, K, and M stars. In such cases, we have adopted the methodology of \citet{segura2005} to manipulate the stellar spectra. From \citet{segura2005}, we normalize and scale the spectrum of AD Leonis (AD Leo), a well-studied, active M3.5Ve star (a star that periodically/episodically flares from a quiescent state), as well as $\epsilon$ Eridani, a K1V star, and $\sigma$ Bo\"otis, an F2V star. Additionally, we have prepared the spectra of the M star GJ 876 from the shortwave observations of the MUSCLES observations \citep{france2013} coupled to a NextGen model for a 3,200 K model star. These spectra, along with the absorption cross-sections for important atmospheric constituents, are shown in \reffig{specx}. Additional details about each star can be found in \reftable{stars}. 

\begin{deluxetable}{cccccccc}
\tabletypesize{\footnotesize}
\tablewidth{0pt}
\tablecaption{Stellar properties \label{stars}}
 \tablehead{
 \colhead{Star} & \colhead{Spectral} & \colhead{T$_{eff}$} & \colhead{Luminosity} & \colhead{Stellar} & \colhead{Distance} & \colhead{Semi-major} & \colhead{Ly-$\alpha$ at 1 AU} \\
 \colhead{} & \colhead{type} & \colhead{(K)} & \colhead{L$_{\odot}$} & \colhead{Radius [R$_{\odot}$]} & \colhead{(pc)} & \colhead{axis (AU)} & \colhead{(ergs cm$^{-2}$ s$^{-1}$)}
 }
 \startdata
 $\sigma$ Bo\"otis & F4V\tablenotemark{1} & 6435\tablenotemark{1} & 3.15\tablenotemark{1} & 1.431\tablenotemark{1} & 15.8\tablenotemark{2} & 2.31 & 79.7\tablenotemark{3} \\
 Sun & G2V & 5780 & 1 & 1 & -- & 1.3 & 8.17 \\
 $\epsilon$ Eridani & K1V\tablenotemark{4} & 5039\tablenotemark{5} & 0.32\tablenotemark{5} & 0.735\tablenotemark{5} & 3.2\tablenotemark{4} & 0.76 & 47.1\tablenotemark{4} \\
 AD Leonis & M3.5Ve\tablenotemark{4} & 3390\tablenotemark{6} & 0.023\tablenotemark{9} & 0.39\tablenotemark{7} & 4.7\tablenotemark{9} & 0.21 & 7.05\tablenotemark{9} \\ 
 GJ 876 & M4V\tablenotemark{8} & 3129\tablenotemark{8} & 0.012\tablenotemark{8} & 0.38\tablenotemark{8} & 4.7\tablenotemark{8} & 0.15 & 0.414\tablenotemark{9} \\ 
\enddata
\tablecomments{
\tablenotemark{1}{\citet{boyajian2013}}; \tablenotemark{2}{\citet{janson2013}}; \tablenotemark{3}{\citet{landsman1993}}; \tablenotemark{4}{\citet{linsky2013}}; \tablenotemark{5}{\citet{baines2012}}; \tablenotemark{6}{\citet{rojas2012}}; \tablenotemark{7}{\citet{reiners2009}}; \tablenotemark{8}{\citet{vonbraun2014}}; \tablenotemark{9}{\citet{france2013}}.
}
\end{deluxetable}


\subsection{Spectral models}
\label{spectral_model}

To generate synthetic direct imaging spectra of planets with the atmospheres modeled in this study, we used the Spectral Mapping Atmospheric Radiative Transfer Model (SMART), developed by D. Crisp \citep{meadows1996,crisp1997}. SMART has been well-validated against spacecraft observations of the Earth \citep{robinson2011,robinson2014} and Venus \citep{arney2014}. An enhanced version of SMART that includes the effects of refraction is used to generate synthetic transit transmission observations of our model atmospheres \citep{misra2014,misra2014a}. SMART's transit transmission capability has been validated against both solar occultation soundings of the Earth's atmosphere \citep{gunson1990} and Earthshine spectral measurements \citep{palle2009}. The absorption cross-sections in the visible to near-infrared wavelength range ($\lambda>$0.4 $\mu$m) for each spectrally active gas are calculated by the SMART routine Line-By-Line ABsorption Coefficients (LBLABC) using the HITRAN 2012 line list database \citep{rothman2013}.  For UV wavelengths ($\lambda<$0.4 $\mu$m), we use the absorption cross sections in \reffig{specx}. To calculate the synthetic transit transmission spectra, we use an index of refraction of $n$ = 1.00031 at STP (consistent with an atmospheric mixture of $\sim$95\% N\2 and $\sim$5\% CO\2) and a planet-star impact parameter of $b$ = 0 (where the planet is assumed to transit directly through the center of the star). For the synthetic direct imaging calculations, we assume the surface is a Lambertian ocean \citep{mclinden1997} and the planet is observed with a solar zenith angle of 60$^{\circ}$, approximating a disk-averaged observation at quadrature phase \citep{segura2005}. We used the stellar parameters given in \reftable{stars}.  The spectral resolution for each case is $\Delta$ν=1 cm$^{-1}$, which is approximately $\Delta\lambda$= 5.8\e{-5} $\mu$m at $\lambda$=0.76 $\mu$m. 

\subsection{Deposition velocities and ocean chemistry}
\label{deps}

Soluble long-lived species are rained out of the troposphere using the parameterization of \citet{giorgi1985}. In some cases, when the ocean is assumed to be saturated with a particular gas, rainout is turned off because it would necessarily be balanced by an upward flux from the ocean surface. Long-lived species may also have a non-zero deposition flux if the ocean is not saturated with that gas. Some detail is given here, as this process can directly affect the atmospheric and global redox balance. Indeed, as we will show, for planets orbiting M stars, the atmospheric O\2 abundance depends critically on the deposition velocities for O\2 and CO, and these quantities, in turn, depend on the chemistry of these gases in solution.

For minor, highly reactive gases and radicals, we assume constant deposition velocities ranging from 0.01-1 cm s$^{-1}$ derived from the modern Earth literature \citep[e.g.,][]{liss1974}. These deposition velocities are limited by the rate of gas transfer within the atmospheric boundary layer. For longer-lived constituents such as O\2 and CO, we use a different approach. Because we are interested in gas transfer to the ocean, rather than surface weathering, we assume that the deposition velocity of such species is controlled by its `piston velocity' within the oceanic boundary layer, (v$_{p}$, [cm s$^{-1}$]) \citep{broecker1982,kharecha2005,domagal2014}. Technically, the oceanic boundary layer resistance adds to the atmospheric resistance, but because the oceanic resistance is much larger, we can simply ignore the atmospheric boundary layer for long-lived species. The piston velocity, in turn, is determined by the vertical gradient in species concentration within a thin film at the top of the ocean and by its molecular diffusivity, K$_{diff}$ (cm$^{2}$ s$^{-1}$), in solution. Mathematically, v$_{p}$ = K$_{diff}$/z$_{film}$, where z$_{film}$ is the (empirically determined) thickness of the film, $\sim$40 $\mu$m. 

To give a specific example, the depositional flux of carbon monoxide into the ocean is calculated from the expression:
\begin{equation}
\Phi(CO) = v_{dep}(CO) n_{CO} = v_{p} \left( \alpha_{CO} p_{CO} - [CO]_{s}\right) \cdot C, \label{dep1}
\end{equation}
Here, $\alpha_{CO}$ (= 1.0\e{-3} mol L$^{-1}$ at 25$^{\circ}$C) is the Henry's Law coefficient for CO, $p_{CO}$ is the partial pressure of CO at the bottom of the atmosphere, $[CO]_{s}$ is the assumed concentration of CO in the surface ocean, and $C$ (= 6.02\e{20} L mol$^{-1}$ cm$^{-3}$) is a unit conversion factor. The piston velocity, $v_{p}$, for CO is 4.8\e{-3} cm s$^{-1}$ \citep{kharecha2005}. The deposition velocity (the values seen in \reftable{species}) is then given by $v_{dep}(CO)$ = $\Phi(CO)/n_{CO}$, where $n_{CO}$ (=2.5\e{19} cm$^{-3}\times f_{CO}$) is the number density of CO at the lowermost grid point. Each calculated deposition velocity bears with it an assumption about the dissolved concentration of that species in the ocean. An upper limit on species deposition velocity is obtained if the dissolved concentration is equal to zero. For CO, at 25$^{\circ}$C, this value is $\sim$1.2\e{-4} cm s$^{-1}$. O\2 has a similar solubility and diffusivity, so under the same conditions its maximum deposition velocity is similar ($\sim$1.4\e{-4} cm s$^{-1}$). 

For species whose concentrations in the ocean are non-zero, calculating the deposition velocity is more complicated. \citet{kharecha2005} calculated abiotic deposition velocities for CO using a 2-box ocean model (surface and deep), along with a limited set of aqueous reactions. They estimated a value of 10$^{-9}$-10$^{-8}$ cm s$^{-1}$, assuming that the largest sink for CO in the ocean was the hydration of CO to formate (CO + OH$^{-} \rightarrow$ HCOO$^{-}$). Here, we expand the aqueous chemistry of Kharecha et al. to include reactions between dissolved O\2, CO, and formate:
\begin{eqnarray}
1/2O_{2} + CO &\rightarrow& CO_{2} \label{aq1}\\
CO + H_{2}O &\rightarrow& CO_{2} + H_{2} \label{aq2}\\
1/2O_{2} + HCOO^{-} &\rightarrow& OH^{-} + CO_{2} \label{aq3}
\end{eqnarray}
These reactions could potentially take place in the ocean or as CO/O\2-bearing water flows through midocean ridges. The first two reactions have been studied in research on catalysis (see \citet{hibbitts2015} for a review); they represent the direct oxidation of CO (\refeq{aq1}; also included in \reffig{scheme}) and the water-gas shift reaction (\refeq{aq2}). An extensive literature survey has not revealed any studies that test these reactions in the absence of catalysts, and we will come back to this issue in the Discussion section. The third suggested reaction (\refeq{aq3}) is the direct oxidation of formate, a process that can be inferred from radiochemical studies \citep[e.g.,][]{balkas1966} and investigations of supercritical water reactions \citep{helling1988}. A moderate amount of atmosphere attenuates X-rays enough to preclude the radiochemical pathway, even for more active stars \citep{cnossen2007}, but this reaction may still occur slowly in the absence of X-rays. These reactions have been incorporated into the model of Kharecha et al., as outlined in the Appendix, and will be used to constrain the deposition velocities for CO and O\2 in the Results section.

\subsection{Balancing the global redox budget}
\label{globalbal}

As discussed in Section \ref{atmobal}, a photochemical model will not satisfy global redox balance automatically; however, one can impose global redox balance by imposing appropriate boundary conditions. In this study, we are primarily interested in calculating upper limits on abiotic O\2. Hence, in \refeq{atmo4}, we neglect the O\2 sinks from oxidative weathering ($\Phi_{OW}$) and from burial of gypsum ($\Phi_{burial}$(CaSO$_{4}$)). In many of our calculations we neglect the O\2 sink due to ferrous iron oxidation ($\Phi_{burial}$(Fe$_{3}$O$_{4}$)), as well, even though this term would arguably be important on an Earth-like exoplanet. We also neglect burial of organic carbon ($\Phi_{burial}$(CH\2O)) and pyrite ($\Phi_{burial}$(FeS\2)). These O\2 sources are important on the modern Earth, but they are both primarily biological. Organic carbon is formed directly by organisms, and pyrite is formed by biologically catalyzed reaction of this organic carbon with dissolved sulfate in the ocean. Oceanic sulfate levels should be low on an abiotic planet, as they were on the Archean Earth prior to the rise of atmospheric O\2 \citep{canfield2000}.

Under these assumptions, the global redox budget can be represented by \refeq{atmo6}. Rainout of oxidants from the atmosphere must equal the rainout of reductants. To balance this equation, we have implemented a scheme similar to that used by \citet{tian2014}. If  $\Phi_{rain}(Ox) <  \Phi_{rain}(Red)$, which is the case for the overwhelming majority of our simulations, the excess hydrogen flowing into the ocean is returned to the atmosphere as H\2. We accomplish this by imposing an upward flux of H\2 at the lower boundary. H\2 is also outgassed from volcanoes, but we simulate this process (and other volcanic outgassing rates) by introducing fictitious chemical source terms within the model troposphere. This procedure allows us to use the lower boundary condition on H\2 to balance the global redox budget.  In the case where $\Phi_{rain}(Ox) >  \Phi_{rain}(Red)$, we impose an increased depositional velocity for H\2. This latter case is rarely observed, though, because the weakly reduced atmospheres studied here consistently produce more soluble reductants than oxidants. All of the results presented here have balanced global redox budgets.

In some cases, we have assumed a non-zero ferrous iron oxidation term, $\Phi_{burial}$(Fe$_{3}$O$_{4}$), in \refeq{atmo4} (the reaction of ferrous iron with O\2 is also included in \reffig{scheme}). Oxidation of dissolved ferrous iron is a quick and readily available O\2 sink, as has been recognized for many years in studies of the early Earth \citep[e.g.,][]{cairns1978,braterman1983,kasting1984}. Biology is not needed to accomplish this process, as it can be catalyzed by near-UV sunlight (ibid.). The deposition of banded iron-formations (BIFs) provides evidence that ferrous iron was abundant in the Archean oceans \citep{holland1973}, and this would probably be true on abiotic exoplanets, as well. That said, most BIF deposition on Earth was probably triggered biologically.

\section{Results}
\label{results}

\subsection{Planets around G stars}
\label{gstar}

\subsubsection{Planets with high H\2 outgassing rates}
\label{ghigh}

We begin by pointing out that our model produces very little ground-level O\2 for typical abiotic Earth conditions, regardless of the CO\2 concentrations in the atmosphere. Here, a typical abiotic Earth would resemble the prebiotic Earth, which is thought to have had more CO\2 than the modern Earth. The adjustment would be caused by the negative feedback between CO\2 and climate \citep{walker1981}. The prebiotic Earth might have had $\sim$80\% N\2 and enough CO\2 ($\sim$20\%) to provide clement conditions. Assumed volcanic outgassing rates for the early Earth are $\sim$3\e{9} cm$^{-2}$ s$^{-1}$ for SO\2, $\sim$3\e{8} cm$^{-2}$ s$^{-1}$ for H\2S, $\sim$3\e{9} cm$^{-2}$ s$^{-1}$ for CH$_{4}$, and $\sim$10$^{10}$ cm$^{-2}$ s$^{-1}$ for H\2 \citep{segura2007}. Under these conditions our model surface O\2 concentrations are $\sim$1\e{-13} PAL, similar to those shown in Fig. 4a of \citet{segura2007}. Outgassing rates within an order of magnitude higher or lower than those quoted here seem reasonable for an Earth-like planet throughout most of its lifetime, and produce similar (very low) surface O\2 concentrations. Here, an `Earth-like' planet is defined as one that has a reduced interior, active volcanism, and a volatile inventory similar to the Earth's.

\subsubsection{Planets with low outgassing rates}
\label{glow}

\begin{figure}
\includegraphics[width=\textwidth]{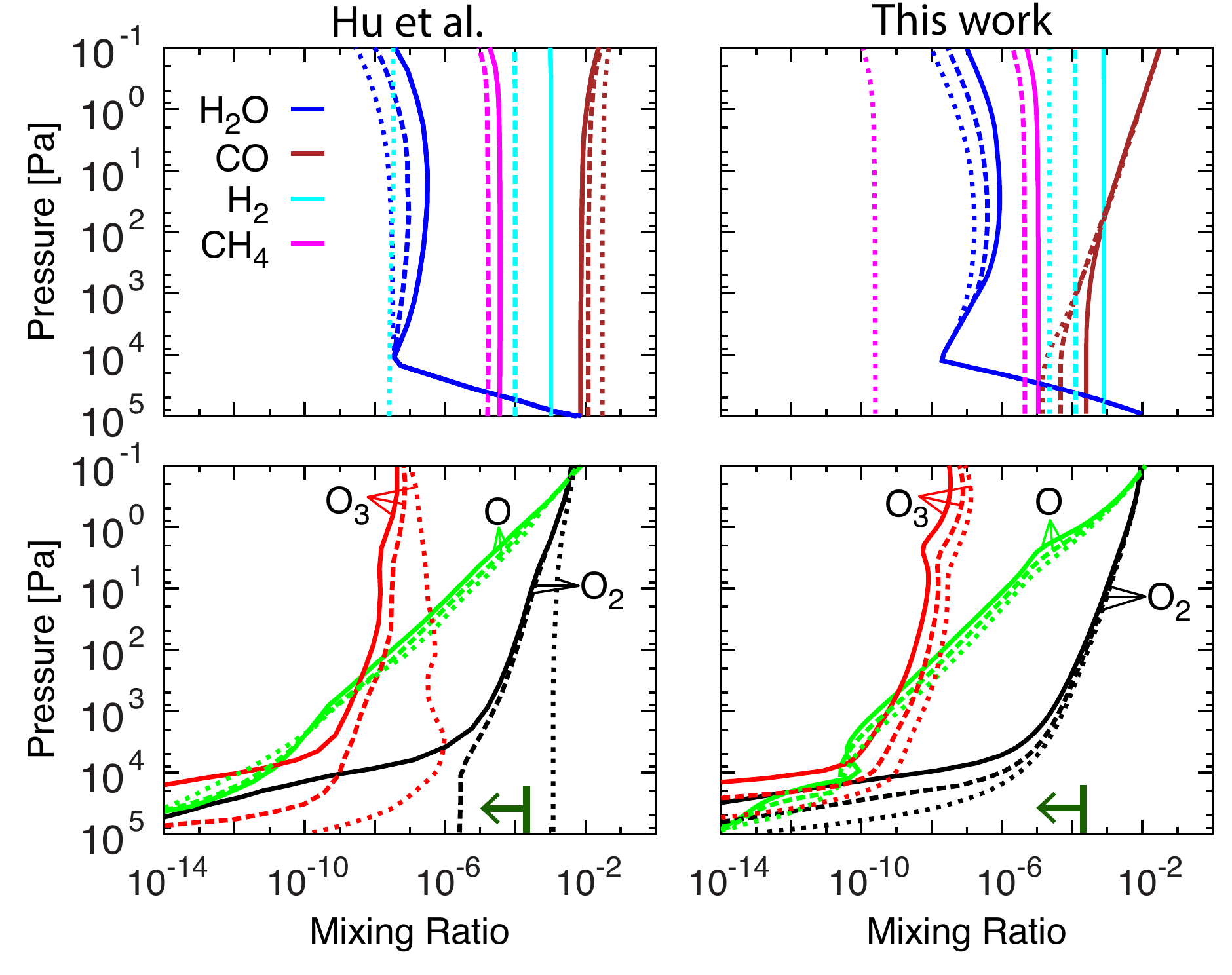}
\caption{Comparative species profiles for the high-CO\2 cases  shown in Fig. 6 of \citet{hu2012} (left panels) and from our work (right panels). A 1-bar, 90 percent CO\2 atmosphere is assumed. Our model balances the global redox budget; Hu et al. balance only the atmospheric redox budget. Solid curves correspond to an H\2 outgassing rate of 3\e{10} cm$^{-2}$ s$^{-1}$; dashed curves are for one-tenth that rate, and the dotted curves correspond to  H\2S outgassing only. The upper limit on Proterozoic O\2 proposed by \citet{planavsky2014} is shown as a green line segment and arrow. \label{hu_comp}}
\end{figure}

We next changed the parameters in the model to try to duplicate false positives that have been published in the recent literature. The first case we looked at was a high-CO\2 atmosphere (90\% CO\2, 10\% N\2) on a planet orbiting a G star like our Sun. \citet{hu2012} simulated such a planet with three different H\2 outgassing rates: 3\e{10}, 3\e{9}, and 0 cm$^{-2}$ s$^{-1}$. Their zero H\2 outgassing case still included a small outgassing flux of H\2S, as discussed further below. The assumed deposition velocities for CO and O\2 were 1\e{-8} cm s$^{-1}$ and zero, respectively. Their results are shown in the two left-hand panels of \reffig{hu_comp}. Surface O\2 was low in the high-outgassing case, but it increased markedly as volcanic outgassing rates decreased. Ultimately, their zero-outgassing case yielded a well-mixed O\2 profile with a mixing ratio of 1.3\e{-3}. As pointed out earlier, this is just slightly above the inferred O\2 mixing ratio for the Proterozoic Earth, so this constitutes a false positive for life by our definition. 

Using the exact same boundary conditions on CO and O\2, we do not generate high surface O\2 concentrations, for reasons that remain unclear (see discussion below). These boundary conditions do not yield a balanced global redox budget, and so we have chosen not to show  them. However, these results differed only slightly from the same calculation with a fully balanced global redox budget and we have shown these latter calculations in the right-hand panels of \reffig{hu_comp}. The global redox budget for these three balanced cases is shown in \reftable{hut}. The surface O\2 mixing ratio in our model never rises above 10$^{-13}$ in any of our simulations; instead, ground-level O\2 mixing ratios are always low, as they are in the high-outgassing case.

\begin{table}
\begin{tabular}{l |c|c|c|}
\cline{2-4}
       & Baseline  & Reduced  & H\2S only  \\ \hline
\multicolumn{1}{ |l| }{H\2 Escape}  & 3.21\e{10}  & 5.10\e{9}  & 9.00\e{8} \\\hline
\multicolumn{1}{ |l| }{Red. Outgassing}  & 3.21\e{10}  & 5.10\e{9}  & 9.00\e{8} \\\hline \hline
\multicolumn{1}{ |l| }{$\Phi_{ox}$}  & 3.57\e{8}  & 6.98\e{8}  & 1.17\e{9} \\ \hline
\multicolumn{1}{ |c| }{SO$_{4}$(aerosol)}  & 3.57\e{8}  & 6.90\e{8}  & 1.11\e{9} \\
\multicolumn{1}{ |c| }{H\2SO$_{4}$}   & 0   & 7\e{6}   & 2\e{7} \\
\multicolumn{1}{ |c| }{H\2O\2}   & 0   & 9\e{5}    & 3\e{7} \\ \hline \hline
\multicolumn{1}{ |l| }{$\Phi_{red}$}  & 4.58\e{9} & 3.30\e{9}  & 1.67\e{9} \\ \hline
\multicolumn{1}{ |c| }{HSO}    & 3.53\e{9} & 2.68\e{9}  & 1.26\e{9} \\
\multicolumn{1}{ |c| }{H\2S}    & 5.1\e{8}  & 3.9\e{8}  & 2.6\e{8} \\
\multicolumn{1}{ |c| }{H\2CO}   & 3.5\e{8}  & 1.3\e{8}  & 8\e{7} \\
\multicolumn{1}{ |c| }{HCO}    & 1.3\e{8}  & 9\e{7}  & 4\e{7} \\
\multicolumn{1}{ |c| }{CO}    & 6\e{7}  & 1\e{7}  & 3\e{6} \\ \hline \hline
\multicolumn{1}{ |l| }{Fe$^{++}$ Flux}  &0    & 0    & 0 \\ \hline
\multicolumn{1}{ |l| }{H\2 Return}   & 4.22\e{9}  & 2.60\e{9}  & 4.96\e{8} \\ \hline \hline
\multicolumn{1}{ |l| }{H\2 Balance}  & 1\e{2}   & 3\e{2}   & 1\e{2} \\ \hline \hline \hline
\multicolumn{1}{ |l| }{O\2 column depth} & 2.1\e{19}  & 3.3\e{19}  & 4.5\e{19} \\ \hline
\multicolumn{1}{ |l| }{O$_{3}$ column depth} & 3.9\e{14}  & 1.5\e{15}  & 4.9\e{15} \\ \hline
\end{tabular}
\caption{Global redox budget for the three cases in the right panels of \reffig{hu_comp}. Baseline: 3\e{10} cm$^{-2}$ s$^{-1}$ H\2, 3\e{8} cm$^{-2}$ s$^{-1}$ CH$_{4}$, 3\e{8} cm$^{-2}$ s$^{-1}$ H\2S; reduced: 3\e{9} cm$^{-2}$ s$^{-1}$ H\2, 3\e{8} cm$^{-2}$ s$^{-1}$ CH$_{4}$, 3\e{8} cm$^{-2}$ s$^{-1}$ H\2S; H\2S only: 3\e{8} cm$^{-2}$ s$^{-1}$ H\2S. All fluxes are in units of cm$^{-2}$ s$^{-1}$. \label{hut}}
\end{table}

Why would Hu et al. predict 10 orders of magnitude more ground-level O\2 than we do for the low-outgassing planet? The answer is not clear. These two photochemical models were developed independently, and both contain many different parameters. The Hu et al. model has a more extensive reaction set and has different rate constants for some reactions, especially those involved in production of CH$_{4}$. It is not obvious, though, why this should cause any significant difference in the results. A model intercomparison study is planned for the coming year to try to resolve this discrepancy.


We can make two observations, though, that go beyond these technical differences in the computer models. First, low-outgassing, Earth-sized planets are probably not physically realistic. Mars is a low-outgassing planet today, but that is because it is only one-tenth of Earth's mass and is volcanically inactive. A true Earth analog would be unlikely to be this dead unless it was extremely old. Second, a planet that was totally volcanically inactive, so that it released zero reduced gases at its surface, could have a much larger O\2 concentration than the one found by Hu et al. According to \refeq{atmo4}, if we neglect surface oxidation and sediment burial terms for such a planet, the volcanic outgassing rate must equal the hydrogen loss rate to space. A zero-outgassing planet must therefore have a zero hydrogen escape rate if its atmosphere is in redox steady state. But Earth-like planets, by definition, are considered to have liquid water on their surfaces and some water vapor in their atmospheres. This alone is enough to create a finite hydrogen escape rate, leaving behind the oxygen from the water. Thus, O\2 should actually accumulate indefinitely on such a planet. One reason it does not in the Hu et al. model, and in our own calculation shown in \reffig{hu_comp}, is that both calculations assumed a finite outgassing rate of H\2S, even while other outgassing rates were set to zero. H\2S can be converted photochemically to H\2 according to the stoichiometry: H\2S + 2 H\2O $\rightarrow$ SO\2 + 3 H\2. So, neither model actually represents a case of zero outgassing of total hydrogen.

The atmospheric redox budget in both models is influenced by rainout and surface deposition of oxidants and reductants at the atmosphere-ocean interface. These processes are particularly important in the Hu et al. model because of the large amount of O\2 present in the lower atmosphere. To first order, surface deposition of the oxidants O$_{3}$, HO\2, and H\2O\2 in their model is balanced by deposition of the reductant CO. (They compute this balance in units of H, instead of H\2, but the implications are the same.) So, if one were to rethink their model in terms of global redox balance, this implies that CO must be reacting with these oxidants in solution within the ocean. That, by itself, is not physically implausible. But O\2 in their model is given a zero deposition velocity at the surface, for reasons that are not explained. What if O\2 itself reacts with CO (or formate) in solution? Its deposition velocity would then be non-zero, and its ground-level mixing ratio would presumably be much lower. As we discuss further below, the rate of aqueous reaction of CO and O\2 is a key parameter in deciding whether false positives are physically realistic. Our own model does not generate a false positive in this situation, so we will not spend further time analyzing it. For now, we simply use this example to illustrate why processes happening below the bottom of one's photochemical model must be taken into account.

\subsection{Planets around M stars}
\label{mstar}

A second possible false positive could occur on planets orbiting M stars. We already dismissed such planets as being habitable in the Introduction, because three different studies have predicted that these planets would lose most or all of their water during the pre-main-sequence phase. That type of false positive, we argued, would be easy to identify because of the lack of H\2O absorption. But let us imagine an M-star planet that either started off with much more water than Earth \citep[e.g.,][]{luger2015a} and sufficient sinks to remove the O\2 that built up from its pre-main sequence time, or a planet that migrated inward to the habitable zone at some time after the parent star had evolved onto the main sequence. Could such a planet accumulate a significant amount of O\2?

This type of planet has been modeled by \citet{tian2014} and \citet{domagal2014}, with conflicting results. Both groups studied a planet orbiting the M star, GJ 876. Tian et al. predict that a planet with 5\% CO\2 in its atmosphere and Earth-like volcanic outgassing rates could reach an O\2 surface mixing ratio of a little more than 2\e{-3} (see their Fig. 3b). This, by our definition, is a false positive. A similar planet orbiting the Sun would have a surface O\2 mixing ratio of more like 10$^{-13}$. (This value is not actually shown on their Fig. 3a, but it can be inferred from our own model, which gives similar results.) The Tian et al. results for the M-star planet are strongly influenced by their lower boundary conditions, as discussed further below. Curiously, when Domagal-Goldman et al. model the same M-star planet, using identical boundary conditions, they find 10$^{4}$ times less O\2. When they use their own lower boundary conditions to find the maximum abiotic O\2 production, O\2 increases by less than a factor of 10, leaving it still 10$^{3}$ times lower than the Tian et al. model. Clearly, lower boundary conditions are important, but other detailed differences between these two different versions of the same parent photochemical model may also be influencing the results.

\begin{figure}
\includegraphics[width=\textwidth]{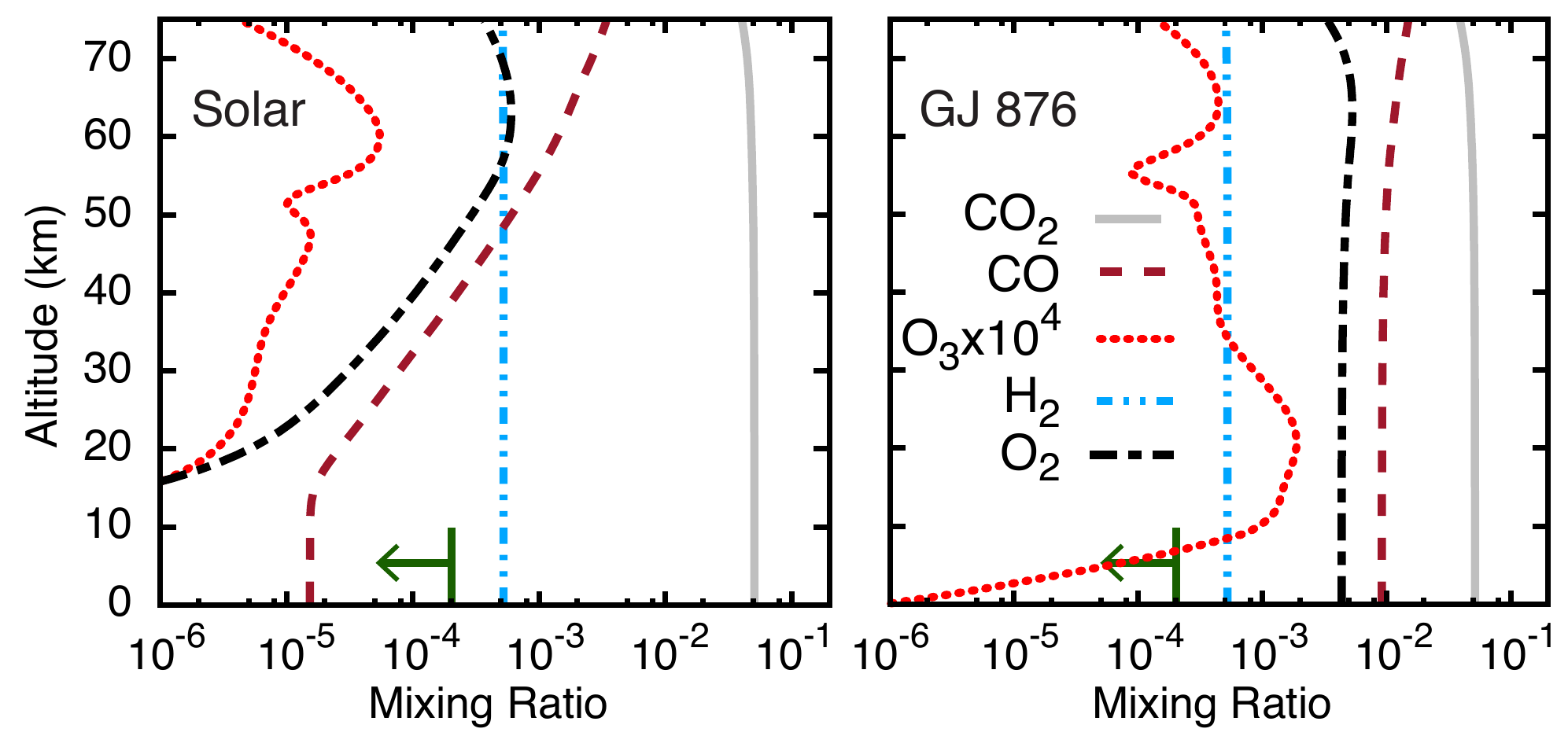}
\caption{Results from our model for comparison with Figure 3 from \citet{tian2014}. Left-hand panel is for the Sun; right-hand panel is for the M star, GJ 876. We used their lower boundary conditions, outgassing rates, and ferrous iron fluxes (see text). We have modified the distance to the planet orbiting GJ 876 to be consistent with other results within this paper. The Planavsky et al. upper limit on Proterozoic O\2 is shown, as in \reffig{hu_comp}. \label{tian_comp}}
\end{figure}

We have qualitatively duplicated the results of Tian et al. in our \reffig{tian_comp}. Our model predicts surface CO and O\2 mixing ratios of $\sim$7\e{-3} and $\sim$3\e{-3}. The slight difference in CO and O\2 mixing ratios compared to their results is a function of the assumed distance to the parent star: because the abiotic Earth modeled by Tian et al. is further from GJ 876, photolysis rates are lower, resulting in less CO and O\2. We note also that Tian et al. have either an implied methane flux or a different set of rates than those reported here; the calculated methane mixing ratio in our model (with no methane outgassing) is about 2 orders of magnitude smaller than theirs for the solar case and is completely negligible for the GJ 876 case (off the scale to the left for both stars in \reffig{tian_comp}). Aside from this, our calculated CO and O\2 surface mixing ratios are quite close to theirs for both the G-star planet and the M-star planet. This agreement may be caused, in part, by the fact that we balanced the global redox budget in the same way that they did -- by adding an upward flux of H\2 from the ocean when rainout of reductants exceeded rainout of oxidants. 


That said, the agreement between our calculations and those of Tian et al. does not imply that the results are realistic. The calculated O\2 mixing ratios for the M-star planet are strongly influenced by the assumed lower boundary conditions which, in this case, are somewhat arbitrary. Following Tian et al., we used an Earth-like rainout rate and deposition velocities of 10$^{-6}$ cm s$^{-1}$ for both CO and O\2. The value for CO is approximately the logarithmic mean of its maximum deposition velocity of 1.2\e{-4} cm s$^{-1}$ and the abiotic deposition velocity of $\sim$10$^{-8}$ cm s$^{-1}$ calculated by \citet{kharecha2005} (see Sect. \ref{modelparam}). That is not a particularly strong justification for choosing this value. The O\2 deposition velocity was set to the same value, again for no clearly identifiable reason. These deposition velocity choices have little effect on the G-star calculation because the surface concentrations of both O\2 and CO are kept low by photochemistry (\reffig{tian_comp}). But for the M-star case, they give rise to large downward fluxes of both species. In our model, the surface O\2 mixing ratio is 3.2\e{-3} and the surface CO mixing ratio is 6.6\e{-3} (\reffig{tian_comp}). The total number density in the lowermost model layer is $\sim$2.5\e{19}, so the downward deposition flux, $\Phi_{dep}^{i} = v_{dep}^{i}n_{i} = v_{dep}^{i}f_{i}n_{tot}$, is 8.0\e{10} cm$^{-2}$ s$^{-1}$ for O\2 and 1.7\e{11} cm$^{-2}$ s$^{-1}$ for CO. The flux of CO is almost exactly equal to twice that of O\2, which is not surprising, given that they are produced in this same ratio by CO\2 photolysis followed by reaction of O with itself to produce O\2. This also implies that these gases must be reacting within the ocean according to: 2 CO + O\2 $\rightarrow$ 2 CO\2. As can be seen from \reftable{tiant}, the CO and O\2 deposition fluxes dominate the atmospheric redox budget in this case, so the answer that one gets is very much a function of the assumed lower boundary conditions on O\2 and CO.

\begin{table}
\begin{tabular}{l |c|c|}
\cline{2-3}
& Sun & GJ 876 \\ \hline
\multicolumn{1}{ |l| }{H\2 Escape} 		& 1.4\e{10} 	& 1.4\e{10} \\\hline
\multicolumn{1}{ |l| }{H\2 Outgassing} 	& 10$^{10}$ 	& 10$^{10}$ \\\hline \hline
\multicolumn{1}{ |l| }{$\Phi_{ox}$} 		& 1.52\e{8} 	& 2.36\e{11} \\ \hline
\multicolumn{1}{ |c| }{O\2} 				& 0 			& 2.34\e{11} \\
\multicolumn{1}{ |c| }{HO\2} 			& 2\e{6} 		& 9\e{6} \\
\multicolumn{1}{ |c| }{H\2O\2} 			& 1.50\e{8} 	& 2\e{9} \\ \hline \hline
\multicolumn{1}{ |l| }{$\Phi_{red}$} 		& 6.45\e{9} 	& 2.38\e{11} \\ \hline
\multicolumn{1}{ |c| }{CO} 				& 4.0\e{8} 		& 2.38\e{11} \\
\multicolumn{1}{ |c| }{HCO} 			& 2.3\e{8} 		& 0 \\
\multicolumn{1}{ |c| }{H\2CO} 			& 5.82\e{9} 	& 0 \\ \hline \hline
\multicolumn{1}{ |l| }{Fe$^{++}$ Flux} 	&4\e{9} 		& 4\e{9} \\ \hline
\multicolumn{1}{ |l| }{H\2 Return} 		& 1.03\e{10} 	& 6.16\e{9} \\ \hline \hline
\multicolumn{1}{ |l| }{H\2 Balance} 		& 3\e{1} 		& 2\e{2} \\ \hline \hline \hline
\multicolumn{1}{ |l| }{O\2 column depth} 	& 1.6\e{19} 	& 9.3\e{22} \\ \hline
\multicolumn{1}{ |l| }{O$_{3}$ column depth} & 5.4\e{14} 	& 8.4\e{17} \\\hline
\end{tabular}
\caption{Global redox budget for \reffig{tian_comp}. All fluxes are in units of cm$^{-2}$ s$^{-1}$. \label{tiant}}
\end{table}

Let us consider what these chosen deposition velocities imply about ocean chemistry. According to \refeq{dep1}, the fact that the chosen deposition velocity is $\sim$1 percent of the maximum deposition velocity means that the ocean must be 99 percent saturated with both CO and O\2. That, in turn, implies that the rate of reaction of these species is very slow. If the rate of reaction of CO and O\2 in solution was fast, then their deposition velocities should be $\sim$10$^{-4}$ cm s$^{-1}$, and their surface mixing ratios would drop by more than a factor of 200. The reaction rate between dissolved CO and O\2 is discussed further in Section \ref{bestcase}.

\subsubsection{`Worst case' scenarios leading to significant false positives} \label{worstcase}

\begin{figure}
\includegraphics[width=0.8\textwidth]{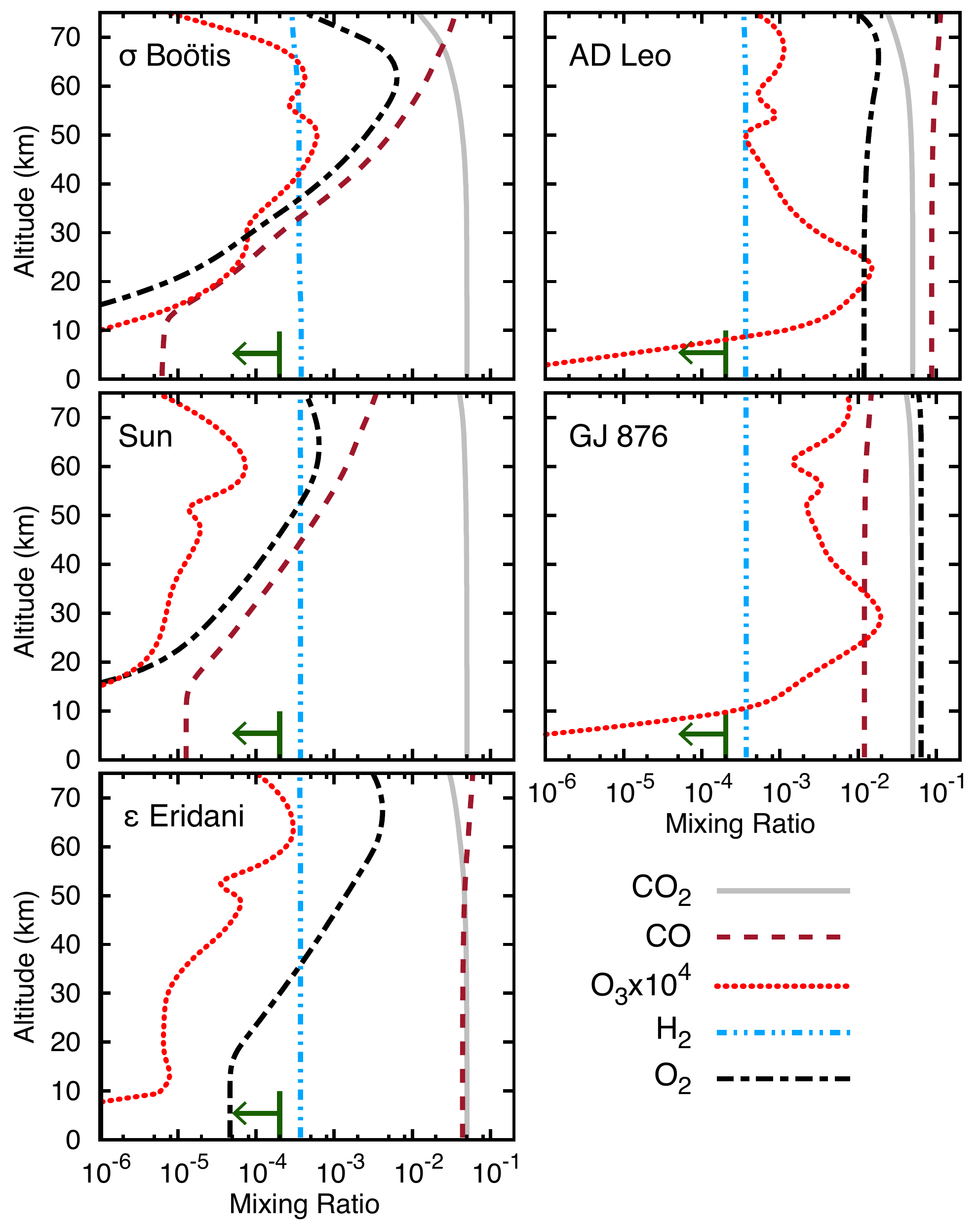}
\caption{Using physically motivated boundary conditions (no sinks for O\2, only abiotic sinks for CO), this figure shows the worst-case (highest O\2) scenario for abiotic planets. We include $\sigma$ Bo\"{o}tis (an F star), AD Leo (an M star), and $\epsilon$ Eridani (a K star) for completeness. Note that the planet around AD Leo is similar to the one around GJ 876 except that it has a higher abundance of CO. Again, the upper limit on Proterozoic O\2 is shown. \label{worst_case}}
\end{figure}

One complication in the methodology being employed is that CO and O\2 are transferred from the atmosphere to the ocean by two different processes -- rainout and surface deposition. For the G-star planet, neither species is produced in appreciable amounts, so both of these fluxes are negligible. For the M-star planet, both gases are much more abundant, and one needs to think carefully about what this assumption implies. For example, a worst-case scenario (\emph{i.e.}, a likely false positive for life) would be if the ocean was completely saturated in O\2. Following \refeq{dep1}, this would imply a zero deposition velocity for O\2. However, the net rainout rate should be zero in this case as well, because any O\2 entering the ocean by way of raindrops would necessarily be balanced by an upward flux of O\2 from the ocean surface. So, the real upper limit on abiotic O\2 comes from setting both rainout and surface deposition equal to zero for O\2. We have illustrated such a case in \reffig{worst_case} for the Sun, GJ 876 (an M star), as well as for $\sigma$ Bo\"{o}tis (an F star), $\epsilon$ Eridani (a K star), and AD Leonis (another M star). \reftable{worstt} details the global hydrogen budget for each case. All of these cases (save for the planet around AD Leo, which is similar to the one around GJ 876) have been used to generate spectra (\reffig{spectra}). For these cases, we kept the deposition velocity for CO at 10$^{-8}$ cm s$^{-1}$, which is equivalent to assuming that formate reacts directly with oxidants in the ocean but that CO does not; thus, the rate-limiting step in CO removal from the surface ocean is hydration to formate. 

\clearpage
\begin{table}
\begin{tabular}{l |c|c|c|c|c||c|}
\cline{2-7}
       								&$\sigma$ Bo\"{o}tis	& Sun		& $\epsilon$ Eridani	& AD Leo  	& GJ 876   	&GJ 876$^{\dag}$\\ \hline
\multicolumn{1}{ |l| }{H\2 Escape}  		& 10$^{10}$  		& 10$^{10}$  	& 10$^{10}$  		& 10$^{10}$ 	& 10$^{10}$ 	& 10$^{10}$ \\\hline
\multicolumn{1}{ |l| }{H\2 Outgassing}  	& 10$^{10}$   		& 10$^{10}$  	& 10$^{10}$  		& 10$^{10}$ 	& 10$^{10}$ 	& 10$^{10}$  \\\hline \hline
\multicolumn{1}{ |l| }{$\Phi_{ox}$}  		& 7.01\e{6}   		& 2.52\e{8}  	& 7.30\e{9}  		& 9.88\e{9} 	& 1.43\e{9} 	& 4.76\e{11}  \\ \hline
\multicolumn{1}{ |c| }{O\2}    			& 0     			& 0    		& 0.    			& 0   			& 0   			& 2.12\e{11}  \\
\multicolumn{1}{ |c| }{HO\2}   			& 3\e{5}    			& 2\e{6}   		& 2.1\e{8}   		& 1\e{7}  		& 5\e{6}  		& 3\e{7}  \\
\multicolumn{1}{ |c| }{H\2O\2}   			& 3.37\e{6}   		& 2.50\e{8}   	& 7.09\e{9} 		& 9.87\e{9} 	&1.42\e{9}  	& 1\e{9} \\ \hline \hline
\multicolumn{1}{ |l| }{$\Phi_{red}$}  		& 6.93\e{10}   		& 5.72\e{9}   	& 1.06\e{10}  		& 2.01\e{10} 	& 2.92\e{9} 	& 4.77\e{11} \\ \hline
\multicolumn{1}{ |c| }{CO}    			& 2\e{6}    			& 3\e{6}    		& 1.06\e{10}  		& 2.01\e{10} 	& 2.92\e{9} 	& 4.77\e{11} \\
\multicolumn{1}{ |c| }{HCO}    			& 1.2\e{9}    		& 2.2\e{8}  	& 0    			& 0   			& 0   			& 0   \\
\multicolumn{1}{ |c| }{H\2CO}   			& 6.81\e{10}   		& 5.49\e{9}  	& 0    			& 0   			& 0   			& 0   \\ \hline \hline
\multicolumn{1}{ |l| }{H\2 Return}   		& 6.93\e{10}   		& 5.47\e{9}   	& 3.30\e{9}  		& 1.03\e{10} 	& 1.49\e{9} 	& 1.00\e{9} \\ \hline \hline
\multicolumn{1}{ |l| }{H\2 Balance}  		& -5\e{2}  			& 5\e{1}    		& 4\e{2}	   		& -2\e{3}  		& -1\e{3}  		& 8\e{1}  \\ \hline \hline \hline
\multicolumn{1}{ |l| }{O\2 column depth} 	& 6.4\e{19}  		& 1.8\e{19}   	& 1.0\e{21}  		& 2.5\e{23} 	& 1.4\e{24} 	& 1.5\e{21} \\ \hline
\multicolumn{1}{ |l| }{O$_{3}$ column depth} & 8.4\e{15}   		& 6.8\e{14}   	& 4.3\e{15}  		& 2.9\e{18} 	& 1.3\e{18} 	& 1.1\e{17} \\\hline
\end{tabular}
\caption{Global redox budget for \reffig{worst_case}, as well as for \reffig{best_case} (denoted with the $^{\dag}$). All fluxes are in units of cm$^{-2}$ s$^{-1}$, and the column depths are in units of cm$^{-2}$. \label{worstt}}
\end{table}
\clearpage

\subsubsection{Scenarios involving surface sinks for O\2} \label{bestcase}

O\2 has several potential sinks on an abiotic world, which we can cast in the form of mass balance equations, following \citet{kharecha2005}. (Our full set of equations is listed in the Appendix.) It is possible that high-temperature supercritical water (for example, as part of hydrothermal circulation) could provide a medium in which CO and O\2 can recombine. Hydrothermal circulation through the midocean ridges would then act as a sink for both species through their direct reaction with one another \citep{helling1988}. Oxygenated waters circulating through the midocean ridges could also oxidize new oceanic crust, creating an additional sink for O\2. The timescale for circulation of ocean water through the midocean ridges on the modern Earth is $\sim$10 My \citep{isley1995}, and could have been lower on the early Earth if the geothermal heat flux was higher. If this is the only significant surface sink for O\2, the O\2 flux into the surface ocean must equal the flux through the midocean ridges. We calculate a lower limit of $\sim$7\e{-9} cm s$^{-1}$ for the O\2 deposition velocity in this case. If v$_{dep}$(O\2) is set equal to this value, and if v$_{dep}$(CO)=10$^{-8}$ cm s$^{-1}$, the calculated O\2 mixing ratio at the surface is $\sim$2\e{-2}. The corresponding downward flux of O\2 is $\sim$10$^{12}$ mol yr$^{-1}$. So, the midocean ridge sink for O\2 would not prevent the occurrence of a false positive for life.

A second potential surface sink for O\2 is the oxidation of formate in solution. If we assume that formate and O\2 react in solution but that CO and O\2 do not, and if this reaction is the only sink for both species, we can calculate the deposition velocity for O\2 as a function of the partial pressure of CO (because CO controls the amount of dissolved formate). For a case with p$_{CO}\approx1$ bar, v$_{dep}$(O\2)$\approx$10$^{-11}$ cm s$^{-1}$ -- well below the deposition velocity estimated for flow through the midocean ridges. The partial pressure of CO must be many orders of magnitude higher than this for formate to be a large sink for O\2. That is unlikely, because it would require more carbon than is available in the Earth's surficial inventory. O\2, however, can be a substantial sink for formate. This would keep dissolved formate at relatively low concentrations, thereby preventing formate from decomposing back into CO. In this case, an assumed CO deposition velocity of 10$^{-8}$ cm s$^{-1}$ is realistic.

Finally, it is conceivable that CO and O\2 could react directly in solution, even though the rate constant for this reaction in a terrestrial ocean (where metal cations may catalyze the recombination of CO and O\2) is unknown. If this reaction were fast, dissolved CO and O\2 concentrations should both approach zero, and their deposition velocities should both be close to their maximum, piston-velocity-limited value (more details in the Appendix). Fig. \ref{best_case} shows what the atmosphere would look like in the case for the planet orbiting GJ 876. The corresponding redox budget is shown in \reftable{worstt}. We have included a spectrum of this case in \reffig{spectra}. The surface O\2 mixing ratio is $\sim$6\e{-5}, which is a factor of 4 below the Proterozoic O\2 concentration inferred by \citet{planavsky2014}. The rapid recombination of dissolved CO and O\2 would be enough to prevent an M-star false positive for life.

In all of the cases examined so far, we have ignored another potentially important sink for O\2, namely, reaction with dissolved ferrous iron in the ocean. On early Earth, the deposition of banded iron-formations suggests that the deep oceans contained abundant ferrous iron \citep{holland1973}. If this were true on the M-star planets simulated above, some of the O\2 entering the ocean should react with Fe$^{2+}$ rather than with CO, thereby lowering the concentration of dissolved O\2 and increasing its deposition velocity. If the flux of ferrous iron entering the oceans was sufficiently large, the dissolved O\2 concentration should be zero, and the O\2 deposition velocity should approach its maximum value of 1.4\e{-4} cm s$^{-1}$ (but the CO deposition velocity would not, so this differs from the case just discussed). \citet{holland2006} estimated the hydrothermal ferrous iron flux entering the present oceans as 3\e{12} mol yr$^{-1}$, or 1.1\e{10} Fe$^{2+}$ cm$^{-2}$ s$^{-1}$. Assume for now that the iron flux into the early oceans was the same. If the iron is oxidized to magnetite, as in \refeq{atmo4}, then 1/6th this amount of O\2 is consumed: 6 FeO + O\2 $\rightarrow$ 2 Fe$_{3}$O$_{4}$. The maximum O\2 flux that could be taken up by reaction with ferrous iron would be $\sim$2\e{9} cm$^{-2}$ s$^{-1}$, or about 1\% of the calculated downward flux of O\2 in the low-O\2 simulation. So, if Holland's numbers apply to exoplanets, ferrous iron oxidation would be only a minor sink for O\2.

That said, ferrous iron fluxes on the early Earth could have been one to two orders of magnitude higher than Holland's estimate, depending on a number of parameters \citep{kump1992,kasting2013}. If all of the Fe$^{2+}$ reacted with O\2, and if the remainder of the O\2 recombined with CO, then the dissolved O\2 concentration should still approach zero, and CO would still be present in excess, keeping its deposition velocity close to the abiotic value of 10$^{-8}$ cm s$^{-1}$. If CO is deposited at this rate, and if O\2 is deposited at its maximum rate, the atmosphere in our model enters CO runaway, in which the calculated CO concentration continues to increase indefinitely. In reality, the CO concentration would eventually be limited by the finite source of CO\2, and so the runaway would stop at some point; however, a dense, CO-dominated atmosphere might still build up. We have not explored such scenarios explicitly; however, we should point out that this type of atmosphere could potentially be identified from the presence of CO absorption features centered at 2.35 and 4.7 $\mu$m. These measurements would require a space-based telescope with near-IR capabilities extending to at least $\sim$2.5 $\mu$m to capture the closest CO feature. The 4.7 $\mu$m CO feature would be more difficult to observe in direct imaging, since it overlaps with the 4.3 $\mu$m CO\2 band, and it occurs near the minimum of the combined planet-and-star emission. However, both the 2.35 $\mu$m and 4.7 $\mu$m CO bands could be observable with transit transmission spectroscopy.

Just prior to the onset of CO runaway, the atmosphere just discussed closely resembles the low-O\2 atmosphere shown in \reffig{best_case}. At higher CO concentrations, calculated O\2 levels fall. Thus, for all these types of atmospheres, O\2 is below the Proterozoic O\2 estimate of \citet{planavsky2014}.

\subsection{Comparisons between planets orbiting F, G, K, and M stars}

\citet{tian2014} suggested that the high stellar far-ultraviolet to near-ultraviolet (FUV/NUV) ratio was responsible for the buildup of O\2 in their M-star case, while \citet{domagal2014} pointed instead to the total FUV flux as the mechanism for building up O$_{3}$ in their F-star case, as discussed in Section \ref{mstar}. In \reffig{fuvnuv}, we have explored this issue by arbitrarily changing the FUV/NUV flux ratio in the solar spectrum. We did this by decreasing the NUV flux and holding the FUV flux constant (NUV is decreasing to the right in \reffig{fuvnuv}). The break between the FUV and the NUV is taken to be 170 nm. We have also plotted the FUV/NUV ratios for the other stars in our study. This was done while using our `worst-case' scenario (5\% CO\2, no rainout of CO and O\2, v$_{dep}$(CO)=10$^{-8}$ cm s$^{-1}$, v$_{dep}$(O\2)=0. cm s$^{-1}$). As the figure shows, the atmosphere transitions from low O\2 to high O\2 as the FUV/NUV ratio increases. Note that $\epsilon$ Eridani (a K star) and $\sigma$ Bo\"{o}tis (an F star) have nearly the same FUV/NUV ratio, and that this ratio roughly corresponds to the point of transition from low O\2 to high O\2. The K-star planet has substantially more O\2 in its atmosphere than does the F-star planet, suggesting that the total UV flux is also an important factor. This effect appears to go the opposite way from that predicted by Domagal-Goldman et al., as the K star has substantially lower FUV fluxes than the F star. We increased the FUV flux while holding NUV constant, to test this hypothesis (results not shown), and found that higher FUV increased O\2 concentrations only slightly. 


The level of FUV has some impact on this result, as it is responsible for dissociating CO\2, but the buildup of O\2 is controlled principally by the NUV flux. We have not considered the effect of flare events, which has previously been examined for O\2-dominated atmospheres \citep{segura2010}, but such flares would likely only affect the atmosphere on short timescales, rather than the long-timescale steady state of our results. In steady state, Tian et al. suggested that the FUV/NUV ratio controls HO\2 and H\2O\2 photolysis, which they suggest are sources of OH. We would point out, however, that these two species are actually just \emph{reservoirs} for OH and that the source of OH is ultimately water vapor photolysis. Decreased water vapor photolysis, alongside the longer lifetime (and greater size) of the HO\2 and H\2O\2 reservoirs, limits the availability of OH for the recombination of CO and O\2 in these atmospheres. In our models, water vapor photolysis slows by a factor of 400 in the GJ 876 case compared to the solar case, and the size of the HO\2 and H\2O\2 reservoirs are 2-3 orders of magnitude larger. These factors together control the O\2 content in our model atmospheres.

 The results of our calculations are summarized in \reffig{allo2}, which shows vertical O\2 profiles for planets around different types of stars. Solid curves represent `worst-case' scenarios, that is, upper limits on abiotic O\2; the dashed curve for the M-star planet is our calculated O\2 concentration for a planet with an ocean in which O\2 reacts rapidly with CO. Planets with additional surface O\2 sinks would have even lower O\2 concentrations, due to the buildup of CO. For F- and G-star planets, the upper limit on O\2 is always well below the biotic (Proterozoic Earth) limit from \citet{planavsky2014}, so such planets should not exhibit false positives for life, according to our calculations. The upper limit on O\2 for the K-star planet is only marginally below the Planavsky et al. limit, suggesting that caution should be exercised when interpreting spectra from these planets. Or, to say this another way, one needs to carefully consider what the surface sinks for O\2 might be on such a planet. For the M-star planet, the upper limit on O\2 (near 0.1 PAL) is almost 100 times greater than the Planavsky et al. limit, and even our lower O\2 estimate is close to that value. Thus, seeing O\2 by itself on a planet around an M star may tell us little about whether that planet is actually inhabited.  


\section{Discussion}

Thus far, we have focused on whether a calculated O\2 or O$_{3}$ concentration could be higher than those on the Proterozoic Earth. But a second important issue is whether or not we could detect the spectral features of abiotically generated O\2 and O$_{3}$. \reffig{spectra} shows synthetic spectra for the abiotic atmospheres with our `worst-case' boundary conditions, \emph{i.e.}, our highest concentrations of O\2 and O$_{3}$. We find that abiotic ozone (O$_{3}$) could be detectable for all cases except those with the Sun and $\epsilon$ Eridani as host stars. The UV O$_{3}$ feature (the Hartley bands) is, unsurprisingly, most prominent for the GJ 876 `worst-case' scenario ($\sim$6\% O\2). The results for AD Leonis are similar and are not shown. Curiously, the M-star planet atmosphere with the lowest O\2 concentration has deeper O$_{3}$ absorption in its reflectivity spectra than the F-star planet; however the F-star planet produces a stronger O$_{3}$ signature in transit. This discrepancy can be explained by the distribution of O$_{3}$ in both atmospheres. While the low-O\2, M-star planet has an O$_{3}$ column depth an order of magnitude greater than the F-star planet (see \reftable{worstt}), the O$_{3}$ mixing ratio is significantly larger at high altitudes on the F-star planet (see Figures \ref{worst_case} and \ref{best_case}). Because transit transmission observations will be insensitive to the lowest layers in the atmosphere in the UV and short-wavelength visible range due to Rayleigh scattering, the high-altitude O$_{3}$ in the F-star planet's atmosphere would be more easily detectable. In neither of these cases should the detectability of O$_{3}$ in the UV be considered evidence for life.

Our synthetic direct imaging and transit transmission spectra show that the O\2-A band would be potentially detectable for M-star planet atmospheres defined by our `worst-case' boundary conditions. For these cases, the O\2-A band strength would be comparable to that of the modern Earth, as the band strength does not increase linearly with O\2 concentration at high O\2 levels.


\section{Conclusions}

Abiotic O\2 concentrations above $\sim$10$^{-3}$ PAL (the estimated O\2 concentration for the Proterozoic Earth) could constitute a false positive for life, even if they are below the detection threshold of current and planned space and ground-based telescopes. O\2 concentrations this high, or higher, are readily produced on planets that lose large amounts of water. This includes runaway greenhouse planets within the inner edge of the HZ and M-star planets that are devolatilized during their pre-main-sequence evolution, which should be recognizable by the lack of water vapor in their atmospheres, and planets whose atmospheres are deficient in nitrogen and carbon (and where H\2O is the major constituent), which should lack N\2-N\2 absorption features. Frozen planets outside the outer edge of the HZ can also build up O\2-rich atmospheres, but these planets should also have H\2O-poor atmospheres. Thus, if the planet's atmospheric composition is constrained, none of these false positives are likely to lead to significant confusion for future UVOIR exoplanet characterization missions.

 Planets more like the Earth, \emph{i.e.}, with liquid water present on their surfaces and orbiting F- and G-type stars, are also unlikely to build up O\2-rich atmospheres, given reasonable assumptions about volcanic outgassing rates and surface oxidation. Earth-like planets orbiting K- and (especially) M-type stars are a potentially bigger problem if they somehow retain surface water after the pre-main-sequence phase of stellar evolution. Such planets are more prone to accumulate O\2 because of the higher far-UV to near-UV stellar flux ratio, as compared to the Sun. Photochemically produced abiotic O\2 levels that exceed the Proterozoic O\2 estimates ($<10^{-3}$ PAL), such as those predicted by \citet{tian2014}, could constitute a possible false positive for life. However, high O\2 concentrations are precluded for planets with sufficient surface sinks for O\2 (e.g., dissolved ferrous iron in the oceans), or if CO and O\2 react rapidly in solution. Laboratory experiments are needed to better determine the aqueous reaction rate of these gases in the presence of (catalytic) metal ions that would likely be present in planetary oceans. In the meantime, one should not rule out O\2 by itself as a potential biosignature, if caution and context are used appropriately.

\acknowledgments

The authors would like to thank Vikki Meadows, Shawn Domagal-Goldman, and Ravi Kopparapu for the helpful discussions that improved the quality of this paper. The authors would also like to thank the anonymous reviewer for their suggestions.

This work was supported in part by the NASA Astrobiology Institute's Virtual Planetary Laboratory Lead Team, funded through the NASA Astrobiology Institute under solicitation NNH12ZDA002C and Cooperative Agreement Number NNA13AA93A.

This work benefited from the use of advanced computational, storage, and networking infrastructure provided by the Hyak supercomputer system at the University of Washington.

\clearpage

\clearpage

\clearpage

\clearpage

\clearpage

\clearpage

\clearpage

\begin{figure}
\includegraphics[width=\textwidth]{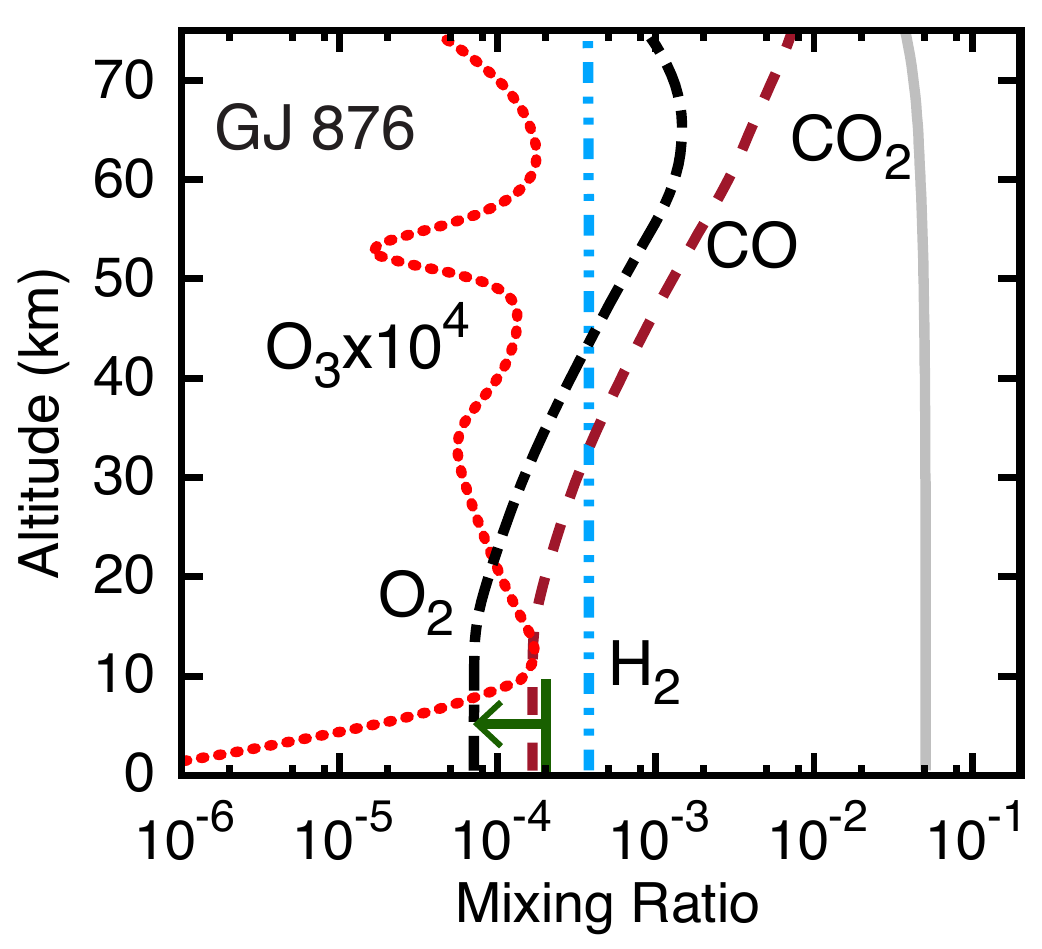}
\caption{Realistic boundary conditions v$_{dep}$(O\2) = 1.4\e{-4} cm s$^{-1}$, v$_{dep}$(CO) = 1.2\e{-4} cm s$^{-1}$, no ferrous iron flux, normal rainout except no rainout of CO or O\2. Only the GJ 876 (M-star) planet is shown, as the change in boundary conditions has little impact on either the solar or $\epsilon$ Eridani planets, and as the effect on the AD Leo planet is nearly identical to that shown here. The upper limit on Proterozoic O\2 is shown. \label{best_case}}
\end{figure}
\clearpage

\begin{figure}
\includegraphics[width=\textwidth]{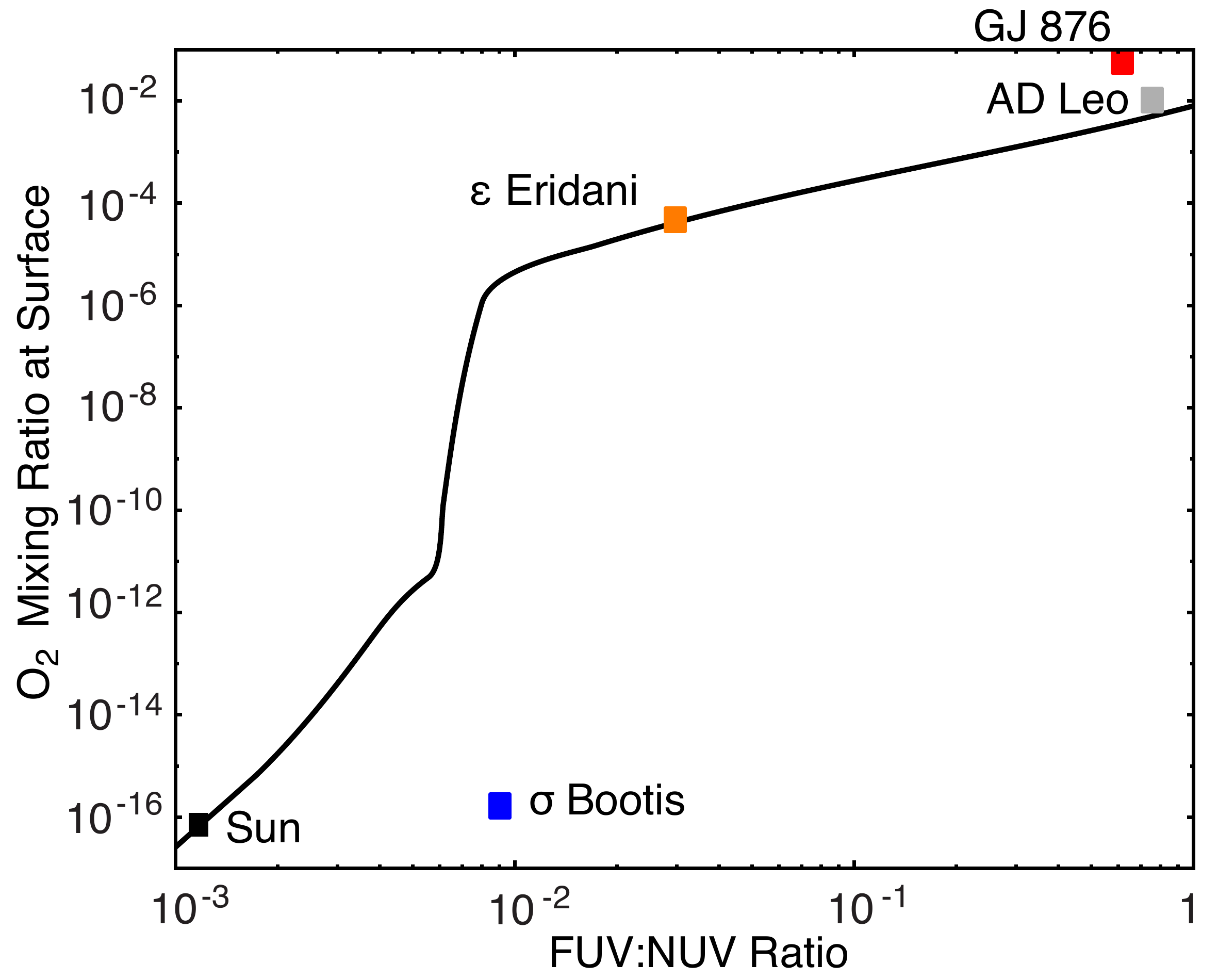}
\caption{Diagram showing the effect of changes in UV flux on the predicted surface O\2 concentration for a planet orbiting the sun. Decreasing the NUV (175 - 320 nm) portion of the solar spectrum while holding the FUV ($<$175 nm) portion constant causes O\2 to increase. The boxes represent the unmodified FUV:NUV ratio for each of the stars in this study.  \label{fuvnuv}}
\end{figure}
\clearpage

\begin{figure}
\includegraphics[width=\textwidth]{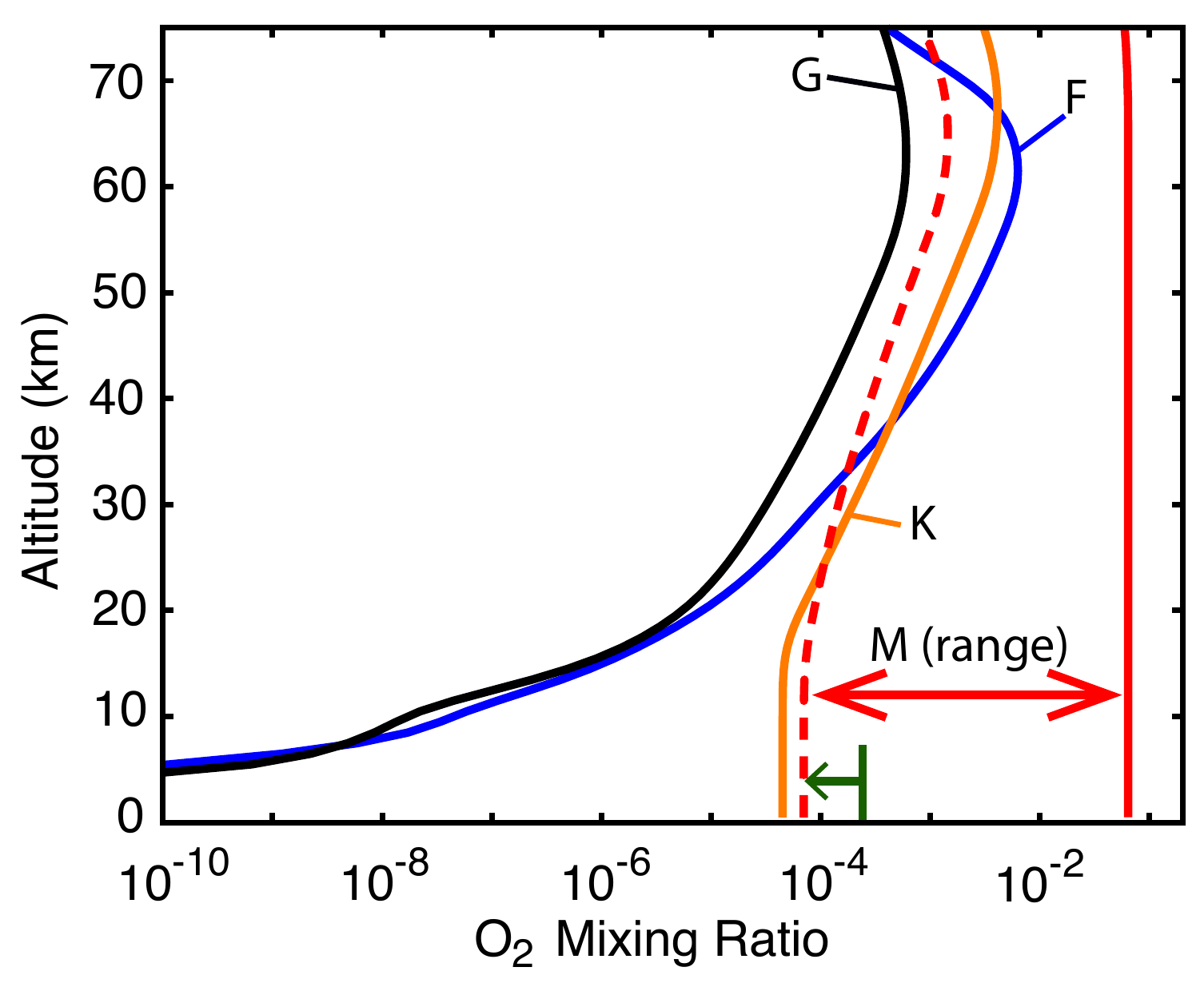}
\caption{Calculated O\2 profiles for abiotic planets orbiting $\sigma$ Bo\"{o}tis (`F'), the Sun (`G'), $\epsilon$ Eridani (`K'), and GJ 876 (`M (range)'). Solid curves represent upper limits on O\2 for cases where volcanic outgassing rates and surface O\2 sinks are small. The dashed curve represents the `low-O\2' case for the M-star planet. The upper limit on Proterozoic O\2 from \citet{planavsky2014} is again shown.
\label{allo2}}
\end{figure}
\clearpage

\begin{figure}
\includegraphics[width=\textwidth]{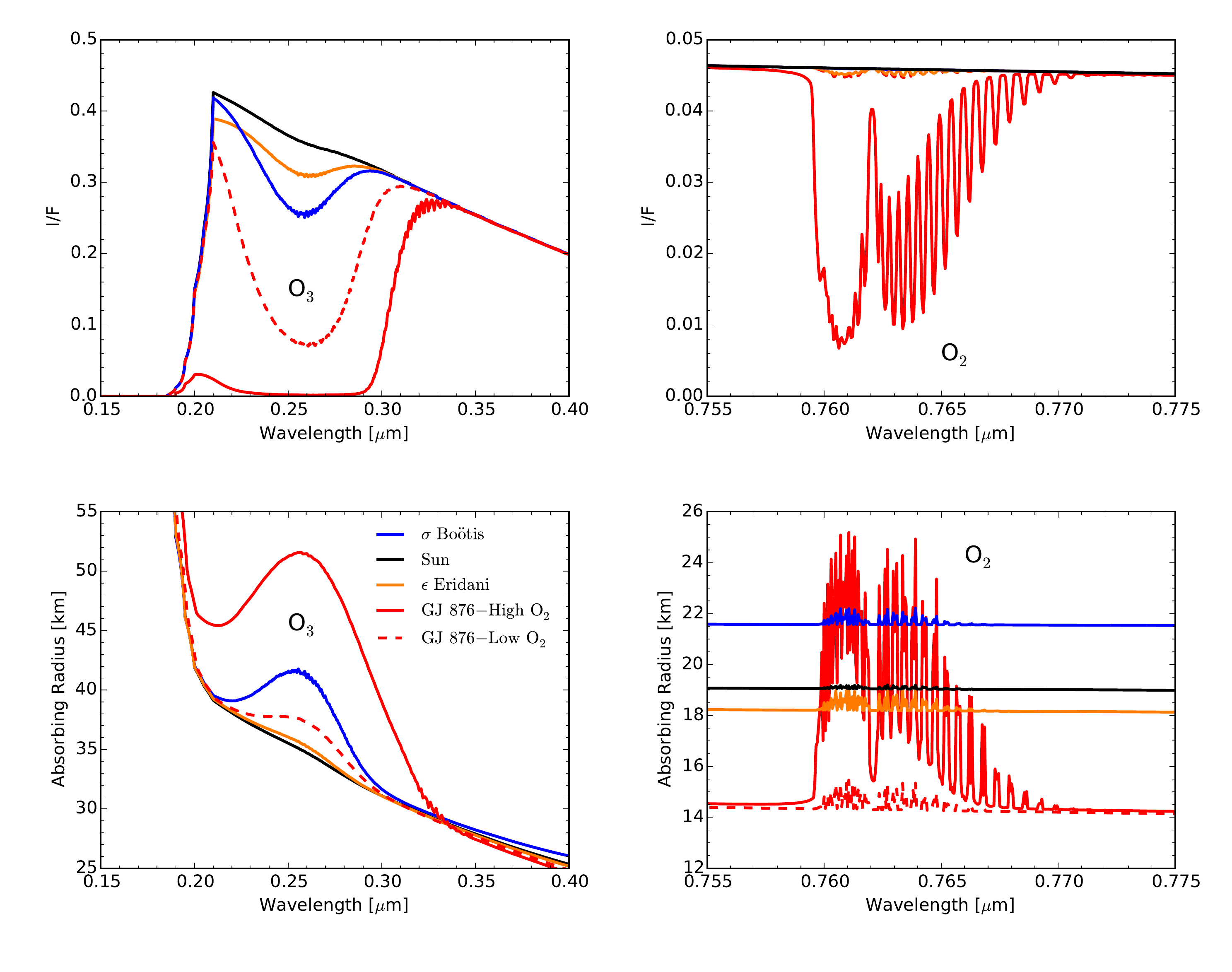}
\caption{Spectra of abiotic (5\% CO\2) atmospheres. Top left: UV (0.15-0.4 $\mu$m) synthetic directing imaging reflectivity spectra for a planet orbiting $\sigma$ Bo\"{o}tis (blue), the Sun (black), $\epsilon$ Eridani (orange), GJ 876 with `low-O\2' boundary conditions (dashed red), and GJ 876 with `worst case' boundary conditions (solid red). The absorption from the O$_{3}$ Hartley bands is evident. Top right: synthetic reflectivity spectrum in the wavelength region of the O\2-A band (0.755-0.775 $\mu$m). Bottom left: synthetic transit transmission spectra in the UV wavelength range in units of the absorbing radius of the atmosphere in kilometers. Bottom right: synthetic transit transmission spectra of the wavelength region containing the O\2-A band. The spectra of AD Leonis are similar to those of the high-O\2  (`worst-case') GJ 876 scenario and not shown here. See section \ref{spectral_model} for a description of the models used to generate these spectra. 
\label{spectra}}
\end{figure}

\clearpage
\appendix

\section{Ocean chemistry and its effect on deposition velocities for O\2 and CO}

As discussed in the text, reactions of O\2 and CO in solution can have a controlling effect on their deposition velocities, and this, in turn, can determine whether atmospheric O\2 concentrations become high enough to constitute a false positive for life. Here, we describe our model for ocean chemistry on an abiotic planet. We acknowledge at the start that several important reaction rate constants are essentially unknown. Thus, instead of providing definitive answers, we focus on what needs to be better constrained in order to calculate realistic O\2 and CO deposition velocities.

We begin from the ocean model of \citet{kharecha2005}. That model included both a surface and a deep ocean, and it did \emph{not} include O\2. We have simplified this model by combining the surface and deep ocean into a single ocean box. This assumption is reasonable because the time scale for deep ocean overturn, about 1,000 years on the present Earth, is shorter than nearly all the chemical reaction time constants in the model. This conclusion could change, of course, as new reaction rates are measured, but for now this simplifying assumption seems justified.

Our ocean model keeps track of the concentrations of three dissolved species: CO, O\2, and HCOO$^{-}$ (formate ion). The suggested chemical reactions involving these species are shown in Section \ref{deps} (eqs. \ref{aq1}-\ref{aq3}). Given this reaction set, we can write a set of mass balance equations for dissolved species concentrations:

\begin{eqnarray}
v_{p}(CO)(\alpha_{CO} p_{CO}  - [CO])\gamma + k_{2}[HCOO^{-}] \frac{z_{s}}{z} \gamma  &&=\nonumber\\ k_{hyd} [OH^{-}] [CO] z \gamma &&+ k_{ox}^{CO} [CO][O_{2}] z \gamma + \Phi_{MOR} [CO] \label{deriv1}\\
v_{p}(O_{2})(\alpha_{O_{2}} p_{O_{2}} - [O_{2}]) \gamma &&=\nonumber\\ k_{ox}^{CO} [CO][O_{2}] z \gamma  + k_{ox}^{HCOO^{-}}[HCOO^{-}] [O_{2}] z \gamma &&+ \Phi_{MOR} [O_{2}] \label{deriv2}\\
k_{hyd} [OH^{-}] [CO] z \gamma = k_{2}[HCOO^{-}] \frac{z_{s}}{z} \gamma &&+ k_{3}[HCOO^{-}] z \gamma + k_{4}[HCOO^{-}] z \gamma \nonumber\\+ k_{ox}^{HCOO^{-}} [HCOO^{-}] [O_{2}] z \gamma &&+ \Phi_{MOR} [HCOO^{-}] \label{deriv3}
\end{eqnarray}
These reactions represent conservation of dissolved O\2, CO, and formate, respectively. The units for these fluxes are mol yr$^{-1}$. The units conversion factor, C, allows us to express dissolved species concentrations in traditional units of mol L$^{-1}$. We assume that all the reactions with unconstrained rates are first order. The known rate constants and other assumed variables (e.g., pH) in our calculations are listed in \reftable{apptable}. Circulation through the midocean ridges is assumed to be a potential sink for all three species.
\begin{deluxetable}{cccc}
\tabletypesize{\footnotesize}
\tablewidth{0pt}
\tablecaption{Model variables and constants \label{apptable}}
 \tablehead{\colhead{Variable} & \colhead{Value} & \colhead{Units} & \colhead{Definition}
 }
\startdata
$v_{p}(CO)$ & 4.8\e{-3} & cm s$^{-1}$ & piston velocity (=D/z) for CO \\
$v_{p}(O_{2})$ & 6.0\e{-3} & cm s$^{-1}$ & piston velocity (=D/z) for O\2 \\
$\alpha_{CO}$ & 8.\e{-4} & mol L$^{-1}$ bar$^{-1}$ & Henry's law coefficient for CO \\
$\alpha_{O_{2}}$ & 1.\e{-3} & mol L$^{-1}$ bar$^{-1}$ & Henry's law coefficient for O\2 \\
$p_{X}$ & -- & bar & surface partial pressure of species $X$ \\
$[X]$ & -- & mol L$^{-1}$ & dissolved concentration of species $X$ \\
$z_{s}$ & 1\e{4} & cm & Surface ocean depth \\
$z$ & 4\e{5} & cm & Global average ocean depth \\
$C$ & 6.02\e{20} & L mol$^{-1}$ cm$^{-3}$ & Units conversion factor \\
$d$ & 0.00374 & L cm$^{-2}$ s$^{-1}$ mol$^{-1}$ & Conversion from mol yr$^{-1}$ to molecules cm$^{-2}$ s$^{-1}$ \\
$\gamma$ & 1.61\e{23} & L s cm$^{-1}$ yr$^{-1}$ & Conversion factor ($=C/d$) \\ 
$k_{hyd}$ & e$^{(-10570/T + 25.6)}$ & L mol$^{-1}$ s$^{-1}$ & Hydration of CO to HCOO$^{-}$ \\
$k_{2}$ & 6.4\e{-5} & s$^{-1}$ & Photodegradation of HCOO$^{-}$ to CO (surface ocean) \\
$k_{3}$ & 2.7\e{-6} & s$^{-1}$ & Acetate formation from HCOO$^{-}$ \\
$k_{4}$ & 8\e{-14} & s$^{-1}$ & Thermal degradation of HCOO$^{-}$ to CO \\ 
$k_{ox}^{CO}$ & \tablenotemark{*} & L mol$^{-1}$ s$^{-1}$ & Direct oxidation of CO (\refeq{aq1}) \\
$\Phi_{MOR}$ & 1.4\e{7} & L yr$^{-1}$ & Ocean flux through the midocean ridges (e.g., \refeq{aq2}) \\
$k_{ox}^{HCOO^{-}}$ & \tablenotemark{*} & L mol$^{-1}$ s$^{-1}$ & Oxidation of HCOO$^{-}$ (\refeq{aq3}) \\
T & 273-298 & K & Range in ocean temperatures (necessary for $k_{hyd}$)\\
pH & 8 & -- & Ocean pH (necessary for [OH$^{-}$]$\approx$10$^{-6}$ mol L$^{-1}$) \\
$p_{surf}$ & 1 & bar & Earth surface pressure
\enddata
\tablenotetext{*}{Unconstrained rate constant} 
\end{deluxetable}
 
With a few assumptions, we can use our model to duplicate the abiotic deposition velocity for CO obtained by Kharecha et al. If CO and O\2 react slowly in solution ($k_{ox}^{CO} \rightarrow 0$), but HCOO$^{-}$ and O\2 react quickly ($k_{ox}^{HCOO^{-}} \rightarrow \infty$), so that HCOO$^{-}$ does not decompose to CO, and if ocean cycling through the midocean ridges is slow, \refeq{deriv1} reduces to:
\begin{equation}
v_{p}(CO)(\alpha_{CO} p_{CO} - [CO]) \gamma = k_{hyd} [OH^{-}] [CO] z \gamma \label{deriv4}
\end{equation}
Rearranging, we can solve for [CO]:
\begin{equation}
[CO] = \alpha_{CO} p_{CO} \frac{1}{1 + k_{hyd} [OH^{-}] z/v_{p}(CO)}
\end{equation}
Substituting this into \refeq{dep1}, we can rearrange terms and solve for v$_{dep}$(CO):
\begin{equation}
v_{dep}(CO) = \frac{v_{p}(CO) \alpha_{CO} C}{n_{air}} \left(1 - \frac{1}{1 + k_{hyd} [OH^{-}] z/v_{p}(CO)}\right)
\end{equation}

Here, we have assumed that pCO/nCO = 1/n$_{air}$, consistent with a 1-bar surface pressure, as well as assuming that the surface ocean resembles that of the modern Earth (pH$\approx$8, T=298 K). For an Earth-like planet, v$_{dep}$(CO)$\approx10^{-8}$ cm s$^{-1}$. This is the deposition velocity for CO if its fastest sink is hydration to formate, and if formate itself is relatively short-lived so that it does not decompose back into CO. If formate is long-lived, the CO deposition velocity would be controlled by acetate formation, and its calculated value would be $\sim$4\e{-10} cm s$^{-1}$. If dissolved CO reacts directly with O\2 by \refeq{aq1}, then the CO deposition velocity could be considerably larger, up to its piston-velocity-limited value of 1.2\e{-4} cm s$^{-1}$. These calculated rates are sensitive to the assumed surface temperature, and are a factor of $\sim$4 higher (lower) for a 15 $^{\circ}$C higher (lower) surface temperature. Looking to eqs. \ref{deriv1}-\ref{deriv3}, we can use this same process to calculate the deposition velocity for other potential sinks for O\2 and CO, which we will detail below. The results of these calculations have been included in Section \ref{bestcase}.

Dissolved O\2 and CO have three potential sinks: 1) flow through the midocean ridges, during which O\2 and CO could react with each other or with the rocks, 2) reaction between O\2 and formate in solution (in which CO is assumed to have hydrated to formate), and 3) direct reaction between O\2 and CO in solution to regenerate CO\2. We will consider each of these in isolation, which is appropriate if one particular sink is dominant. 

\subsection{O\2 and the Midocean ridges}
If the only sink for O\2 is circulation through the midocean ridges then we can write: 
\begin{equation}
v_{p}(O_{2})(\alpha_{O_{2}}p_{O_{2}} - [O_{2}])\gamma = \Phi_{MOR}[O_{2}].
\end{equation}
Setting these two terms equal to each other, solving for [O\2], and incorporating an expression for the downward flux of O\2 analogous to \refeq{dep1} in the main text, we can determine a lower limit of v$_{dep}$(O\2)$\approx$7\e{-9} cm s$^{-1}$. The sink for CO from this same process would result in v$_{dep}$(CO)$\approx$6\e{-9} cm s$^{-1}$, assuming that CO either reacts with O\2 at high temperatures or undergoes the water-gas shift reaction (\refeq{aq2}). The latter case provides a physical mechanism to support our method for balancing the redox budget with a return flux of H\2, as the calculated deposition velocity from this process is comparable to the assumed abiotic deposition velocity for CO.

\subsection{O\2 and formate}
If formate and O\2 react quickly in solution (\emph{i.e.}, $k_{ox}^{HCOO^{-}}>>1$ L mol$^{-1}$ s$^{-1}$), and if we further assume that this reaction dominates both the O\2 and formate systems, we can set the terms from eqs. \ref{deriv1} and \ref{deriv2} equal to one another:
\begin{equation}
v_{p}(O_{2})(\alpha_{O_{2}}p_{O_{2}} - [O_{2}])\gamma  = k_{ox}^{HCOO^{-}} [HCOO^{-}][O_{2}]z\gamma. 
\end{equation}
To determine p$_{CO}$ and p$_{O_{2}}$ simultaneously, the system would need to evolve forward in time. We can, however, solve for the deposition velocity as a function of p$_{CO}$, with the deposition velocity represented as a fraction of the maximum deposition velocity (thus reducing the system to one variable). For example, to achieve 90\% of the maximum O\2 deposition velocity through reactions with formate, p$_{CO}>10^{8}$ bars, assuming the ocean is fully saturated in CO ([CO] = $\alpha_{CO}$p$_{CO}$). If the ocean were undersaturated with respect to CO, or if the reaction rate between O\2 and formate were slower, p$_{CO}$ would have to be even larger. For our worst-case scenario with p$_{CO}\approx0.01$ bar, v$_{dep}$(O\2)$\approx$10$^{-13}$ cm s$^{-1}$. Increasing p$_{CO}$ to 1 bar increases the calculated deposition velocity to $\sim10^{-11}$ cm s$^{-1}$.

\subsection{O\2 and CO}
We can calculate the necessary reaction rate for the consumption of all available O\2 via the direct oxidation of CO (\refeq{aq1}). We do this by neglecting all other O\2 sinks and equating the flux of O\2 into the ocean with the oxidation rate of CO:
\begin{equation}
v_{p}(O_{2})(\alpha_{O_{2}}p_{O_{2}} - [O_{2}])\gamma = k_{ox}^{CO}[CO][O_{2}]z\gamma. 
\end{equation}
We can use our low-O\2 scenario for GJ 876 to establish a minimum for [CO] based on the CO flux into the ocean and its lifetime with respect to circulation through the midocean ridges. If v$_{dep}$(O\2) is 90\% of its maximum, we can constrain k$_{ox}^{CO}\ge$4\e{5} L mol$^{-1}$ s$^{-1}$. If the actual value for this reaction is lower than this, or if the reaction is not first-order in either reactant, then v$_{dep}$(O\2) would fall from its maximum; however, a rate even two orders of magnitude slower maintains v$_{dep}$(O\2) at 10\% of its maximum, suggesting that a non-zero reaction rate for the direct oxidation of CO would be sufficient to draw down surface O\2 below the Planavsky et al. limit.

\clearpage
\section{Reaction Table}
\begin{deluxetable}{l|c|c|c}
\renewcommand{\arraystretch}{1.5}
\tabletypesize{\small}
\rotate
\tablecaption{Reactions and rates.}
\tablewidth{0pt}
\tablehead{
\colhead{Rxn \#} & \colhead{Reaction} & \colhead{Rate [cm$^{3}$ s$^{-1}$]} & \colhead{Source} 
}
\startdata
1 	& H\2O + O($^{1}$D) $\rightarrow$ OH + OH   					& 1.63\e{-10}\EXP{60}{T}					&  \tablenotemark{1}   \\
2 	& H\2 + O($^{1}$D) $\rightarrow$ OH + H   					& 1.2\e{-10} 							&  \tablenotemark{1}  \tablenotemark{a} \\
3 	& H\2 + O $\rightarrow$ OH + H   							& 3.44\e{-13}$\cdot$ (T/298 K)$^{2.67} \cdot$ \EXP{-3160}{T} 	&  \tablenotemark{2}   \\
4 	& H\2 + OH $\rightarrow$ H\2O + H   						& 2.8\e{-12}$\cdot$ \EXP{-1800}{T} 			&  \tablenotemark{1}   \\
5 	& H + O$_{3}$ $\rightarrow$ OH + O\2   						& 1.4\e{-10}$\cdot$ \EXP{-470}{T} 			&  \tablenotemark{1}   \\\hline
6 	& H + O\2 + M $\rightarrow$ HO\2 + M   						& \tbdy{5.7\e{-32}}{-1.6}{7.5\e{-11}}{0} 		&  \tablenotemark{1} \tablenotemark{b}   \\
7 	& H + HO\2 $\rightarrow$ H\2 + O\2   						& 6.9\e{-12} 							&  \tablenotemark{1}   \\
8 	& H + HO\2 $\rightarrow$ H\2O + O   						& 1.6\e{-12} 							&  \tablenotemark{1}   \\
9 	& H + HO\2 $\rightarrow$ OH + OH   						& 7.2\e{-11} 							&  \tablenotemark{1}   \\
10 	& OH + O $\rightarrow$ H + O\2   							& 1.8\e{-11}$\cdot$ \EXP{180}{T} 			&  \tablenotemark{1}   \\\hline
11 	& OH + HO\2 $\rightarrow$ H\2O + O\2   						& 4.8\e{-11}$\cdot$ \EXP{250}{T} 			&  \tablenotemark{1}   \\
12 	& OH + O$_{3}$ $\rightarrow$ HO\2 + O\2   					& 1.7\e{-12}$\cdot$ \EXP{-940}{T} 			&  \tablenotemark{1}   \\
13 	& HO\2 + O $\rightarrow$ OH + O\2   						& 3.0\e{-11}$\cdot$ \EXP{200}{T} 			&  \tablenotemark{1}   \\
14 	& HO\2 + O$_{3}$ $\rightarrow$ OH + O\2 + O\2 				& 1.0\e{-14}$\cdot$ \EXP{-490}{T} 			&  \tablenotemark{1}   \\
15 	& HO\2 + HO\2 $\rightarrow$ H\2O\2 + O\2   					& 3.0\e{-13}$\cdot$ \EXP{460}{T} 			&  \tablenotemark{1}   \\\hline
16 	& H\2O\2 + OH $\rightarrow$ HO\2 + H\2O  			 		& 1.8\e{-12}					 		&  \tablenotemark{1}   \\
17 	& O + O + M $\rightarrow$ O\2 + M   						& 5.21\e{-35}$\cdot$ \EXP{900}{T}$\cdot$[M] 	&  \tablenotemark{3}   \\
18 	& O + O\2 + M $\rightarrow$ O$_{3}$ + M   					& \tbdy{6.0\e{-34}}{-2.4}{1.0\e{-10}}{0} 		&  \tablenotemark{1} \tablenotemark{b}   \\
19 	& O + O$_{3}$ $\rightarrow$ O\2 + O\2   						& 8.0\e{-12}$\cdot$ \EXP{-2060}{T} 			&  \tablenotemark{1}   \\
20 	& OH + OH $\rightarrow$ H\2O + O   						& 1.8\e{-12} 							&  \tablenotemark{1}   \\\hline
21 	& O($^{1}$D) + M $\rightarrow$ O + M   						& 1.8\e{-11}$\cdot$ \EXP{110}{T} 		&  \tablenotemark{1}  \tablenotemark{a} \\
22 	& O($^{1}$D) + O\2 $\rightarrow$ O + O\2   					& 3.3\e{-11}$\cdot$ \EXP{55}{T} 			&  \tablenotemark{1}   \\
23 	& O\2 + h$\nu$ $\rightarrow$ O + O($^{1}$D)   					& 1.5\e{-6}; 2.6\e{-6}					&   \tablenotemark{c} \\
24 	& O\2 + h$\nu$ $\rightarrow$ O + O   						& 2.9\e{-8}; 5.1\e{-9}					&   \tablenotemark{c} \\
25 	& H\2O + h$\nu$ $\rightarrow$ H + OH   						& 4.3\e{-6}; 8.9\e{-6}					&   \tablenotemark{c} \\\hline
26 	& O$_{3}$ + h$\nu$ $\rightarrow$ O\2 + O($^{1}$D)  			& 3.3\e{-3}; 9.1\e{-6}					&   \tablenotemark{c} \\
27 	& O$_{3}$ + h$\nu$ $\rightarrow$ O\2 + O   					& 9.7\e{-4}; 3.9\e{-5}					&   \tablenotemark{c} \\
28 	& H\2O\2 + h$\nu$ $\rightarrow$ OH + OH   					& 5.0\e{-5}; 2.3\e{-7}					&   \tablenotemark{c} \\
29 	& CO\2 + h$\nu$ $\rightarrow$ CO + O   						& 4.4\e{-10}; 7.4\e{-11}				&  \tablenotemark{c}  \\
30 	& CO + OH $\rightarrow$ CO\2 + H   						& \tbdy{1.5\e{-13}/[M]}{0.6}{2.1\e{9}/[M]}{6.1}	&  \tablenotemark{1}  \tablenotemark{a} \\\hline
31 	& CO + O + M $\rightarrow$ CO\2 + M   						& 1.7\e{-33}$\cdot$ \EXP{-1550}{T}$\cdot$ [M] &  \tablenotemark{3}   \\
32 	& H + CO + M $\rightarrow$ HCO + M   						& 2.0\e{-33}$\cdot$ \EXP{-850}{T}$\cdot$ [M] &  \tablenotemark{2}   \\
33 	& H + HCO $\rightarrow$ H\2 + CO   						& 1.83\e{-10} 						&  \tablenotemark{4}   \\
34 	& HCO + HCO $\rightarrow$ H\2CO + CO   					& 4.48\e{-11} 						&  \tablenotemark{4}   \\
35 	& OH + HCO $\rightarrow$ H\2O + CO   						& 1.69\e{-10} 						&  \tablenotemark{2}   \\\hline
36 	& O + HCO $\rightarrow$ H + CO\2   						& 5.0\e{-11} 						&  \tablenotemark{3}   \\
37 	& O + HCO $\rightarrow$ OH + CO   						& 5.0\e{-11} 						&  \tablenotemark{3}   \\
38 	& H\2CO + h$\nu$ $\rightarrow$ H\2 + CO   					& 2.8\e{-5}; 5.3\e{-7}					&  \tablenotemark{c}  \\
39 	& H\2CO + h$\nu$ $\rightarrow$ HCO + H   					& 3.4\e{-5}; 2.4\e{-7}					& \tablenotemark{c}   \\
40 	& HCO + h$\nu$ $\rightarrow$ H + CO   						& 1\e{-2} 							&  \tablenotemark{5}   \\\hline
41 	& H\2CO + H $\rightarrow$ H\2 + HCO   						& 1.44\e{-11}$\cdot$ \EXP{-1744}{T} 	& \tablenotemark{6}   \\
42 	& CO\2 + h$\nu$ $\rightarrow$ CO + O($^{1}$D) 				& 6.5\e{-8}; 1.6\e{-7}					&  \tablenotemark{c} \\
43 	& H + H + M $\rightarrow$ H\2 + M   							& 8.85\e{-33}$\cdot$ T$^{-0.6}$ 		&  \tablenotemark{2}   \\
44 	& HCO + O\2 $\rightarrow$ HO\2 + CO   						& 5.2\e{-12}			 			&  \tablenotemark{1}   \\
45 	& H\2CO + OH $\rightarrow$ H\2O + HCO   					& 5.5\e{-12}$\cdot$\EXP{125}{T}		&  \tablenotemark{1}   \\\hline
46 	& H + OH + M $\rightarrow$ H\2O + M   						& 6.1\e{-26}$\cdot$ T$^{-2}$  $\cdot$ [M]	&  \tablenotemark{3}  \\
47 	& OH + OH + M $\rightarrow$ H\2O\2 + M   					& \tbdy{6.9\e{-31}}{-1}{2.6\e{-11}}{0} 		&  \tablenotemark{1}  \tablenotemark{b}  \\
48 	& H\2CO + O $\rightarrow$ HCO + OH   						& 1.78\e{-11}$\cdot$ (T/298)$^{0.57} \cdot$ \EXP{-1400}{T} 		&  \tablenotemark{2}   \\
49 	& H\2O\2 + O $\rightarrow$ OH + HO\2   						& 1.4\e{-12}$\cdot$ \EXP{-2000}{T} 		&  \tablenotemark{1}   \\
50 	& HO\2 + h$\nu$ $\rightarrow$ OH + O   						& 3.3\e{-4}; 1.1\e{-6}					&  \tablenotemark{c}  \\\hline
51 	& CH$_{4}$ + h$\nu$ $\rightarrow$ $^{1}$CH\2 + H\2 			& 8.2\e{-7}; 2.9\e{-6}					&   \tablenotemark{c} \\
52 	& C\2H$_{6}$ + h$\nu$ $\rightarrow$ $^{3}$CH\2 + $^{3}$CH\2 + H\2 	& 0; 0						&   \tablenotemark{c} \\
53 	& C\2H$_{6}$ + h$\nu$ $\rightarrow$ CH$_{4}$ + $^{1}$CH\2   	& 3.8\e{-7}; 1.3\e{-6}					&   \tablenotemark{c} \\
54 	& HNO\2 + h$\nu$ $\rightarrow$ NO + OH   					& 1.7\e{-3}; 1.7\e{-3} 					& \tablenotemark{c}   \\
55 	& HNO$_{3}$ + h$\nu$ $\rightarrow$ NO\2 + OH   				& 7.2\e{-5}; 1.1\e{-6}					&  \tablenotemark{c}  \\\hline
56 	& NO\2 + h$\nu$ $\rightarrow$ NO + O   						& 3.4\e{-3}; 1.5\e{-4}					&  \tablenotemark{c}  \\
57 	& CH$_{4}$ + OH $\rightarrow$ CH$_{3}$ + H\2O   				& 2.45\e{-12}$\cdot$ \EXP{-1775}{T} 	&  \tablenotemark{1}   \\
58 	& CH$_{4}$ + O($^{1}$D) $\rightarrow$ CH$_{3}$ + OH   			& 1.31\e{-10} 						&  \tablenotemark{1}   \\
59 	& CH$_{4}$ + O($^{1}$D) $\rightarrow$ H\2CO + H\2   			& 9.0\e{-12} 						&  \tablenotemark{1}   \\
60 	& $^{1}$CH\2 + CH$_{4}$ $\rightarrow$ CH$_{3}$ + CH$_{3}$   	& 5.9\e{-11} 						& \tablenotemark{7}   \\\hline
61 	& $^{1}$CH\2 + O\2 $\rightarrow$ HCO + OH   					& 3\e{-11} 							& \tablenotemark{8}   \\
62 	& $^{1}$CH\2 + M $\rightarrow$ $^{3}$CH\2 + M   				& 8.8\e{-12} 						& \tablenotemark{8}   \\
63 	& $^{3}$CH\2 + H\2 $\rightarrow$ CH$_{3}$ + H  				& 5.0\e{-14} 						& \tablenotemark{9}   \\
64 	& $^{3}$CH\2 + CH$_{4}$ $\rightarrow$ CH$_{3}$ + CH$_{3}$  	& 7.1\e{-12}$\cdot$ \EXP{-5051}{T} 		& \tablenotemark{9}   \\
65 	& $^{3}$CH\2 + O\2 $\rightarrow$ HCO + OH   					& 1.5\e{-12} 						& \tablenotemark{10} 	  \\\hline
66 	& CH$_{3}$ + O\2 + M $\rightarrow$ H\2CO + OH   				& \tbdy{4.0\e{-31}}{-3.6}{1.2\e{-12}}{1.1} 	&  \tablenotemark{1}  \tablenotemark{b}  \\
67 	& CH$_{3}$ + OH $\rightarrow$ H\2CO + H\2   					& 9.1\e{-11}$\cdot$ \EXP{-1500}{T}		& \tablenotemark{11}   \\
68 	& CH$_{3}$ + O $\rightarrow$ H\2CO + H   					& 1.4\e{-10} 						&  \tablenotemark{2}   \\
69 	& CH$_{3}$ + O$_{3}$ $\rightarrow$ H\2CO + HO\2   			& 5.4\e{-12}$\cdot$ \EXP{-220}{T} 		&  \tablenotemark{1}   \\
70 	& CH$_{3}$ + CH$_{3}$ + M $\rightarrow$ C\2H$_{6}$     			& \tbdyg{0.381$\cdot$\EXP{-T}{73.2} + 0.61$\cdot$\EXP{-T}{1180}}{8.76\e{-7}$\cdot$T$^{-7.03}\cdot$ \EXP{-1390}{T}$\cdot$[M]}{1.5\e{-7}$\cdot$T$^{-1.18}\cdot$\EXP{-329}{T}}					& \tablenotemark{12} \tablenotemark{d}  \\\hline
71 	& CH$_{3}$ + h$\nu$ $\rightarrow$ $^{1}$CH\2 + H   			& 1.9\e{-4}; 1.1\e{-4}					& \tablenotemark{c}   \\
72 	& CH$_{3}$ + H + M $\rightarrow$ CH$_{4}$ + M   				&  \tbdyg{0.902-(1.03\e{-3}$\cdot$T)}{4.0\e{-29}$\cdot$[M]}{4.7\e{-10}} 					& \tablenotemark{13}  \tablenotemark{d} \\
73 	& CH$_{3}$ + HCO $\rightarrow$ CH$_{4}$ + CO   				& 2.01\e{-10} 						&  \tablenotemark{3}   \\
74 	& CH$_{3}$ + HNO $\rightarrow$ CH$_{4}$ + NO   				& 1.85\e{-11} $\cdot$ (T/298)$^{0.76} \cdot$ \EXP{175}{T} 	& \tablenotemark{14}   \\
75 	& CH$_{3}$ + H\2CO $\rightarrow$ CH$_{4}$ + HCO   			& 1.6\e{-16} $\cdot$ (T/298)$^{6.1} \cdot$ \EXP{-990}{T}		& \tablenotemark{15}   \\\hline
76 	& H + NO + M $\rightarrow$ HNO + M   						& 1.22\e{-33} $\cdot$ (T/298)$^{-1.17} \cdot$ \EXP{-210}{T} &  \tablenotemark{3}  \\
77 	& NO + O$_{3}$ $\rightarrow$ NO\2 + O\2   					& 3.0\e{-12}$\cdot$ \EXP{-1500}{T} 		&  \tablenotemark{1}  \\
78 	& NO + O + M $\rightarrow$ NO\2 + M   						& \tbdy{9.0\e{-32}}{-1.5}{3.0\e{-11}}{0} 	&  \tablenotemark{1}   \tablenotemark{b} \\
79 	& NO + HO\2 $\rightarrow$ NO\2 + OH   						& 3.3\e{-12}$\cdot$ \EXP{270}{T} 		&  \tablenotemark{1}   \\
80 	& NO + OH + M $\rightarrow$ HNO\2 + M   					& \tbdy{7.0\e{-31}}{2.6}{3.6\e{-11}}{0.1} 	&  \tablenotemark{1}  \tablenotemark{b}  \\\hline
81 	& NO\2 + O $\rightarrow$ NO + O\2   						& 5.1\e{-12}$\cdot$ \EXP{210}{T} 		&  \tablenotemark{1}   \\
82 	& NO\2 + OH + M $\rightarrow$ HNO$_{3}$ + M   				& \tbdy{1.8\e{-30}}{3.0}{2.8\e{-11}}{0} 	& \tablenotemark{1}  \tablenotemark{b}  \\
83 	& NO\2 + H $\rightarrow$ NO + OH   						& 4.0\e{-10}$\cdot$ \EXP{-340}{T} 		&  \tablenotemark{16}   \\
84 	& HNO$_{3}$ + OH $\rightarrow$ H\2O + NO\2 + O 				& $\Bigg |$\parbox{8cm}{2.4\e{-14}$\cdot$ \EXP{460}{T} + 6.5\e{-34}$\cdot$\EXP{1335}{T} $\cdot$(1 + 6.5\e{-34}$\cdot$\EXP{1335}{T}/2.7\e{-17}$\cdot$\EXP{2199}{T})}  &  \tablenotemark{1}   \\
85 	& HCO + NO $\rightarrow$ HNO + CO   						& 1.2\e{-10}$\cdot$ T$^{-0.4}$ 			&  \tablenotemark{17}   \\\hline
86 	& HNO + h$\nu$ $\rightarrow$ NO + H   						& As R54 							& \tablenotemark{e}    \\
87 	& H + HNO $\rightarrow$ H\2 + NO   						& 3.01\e{-11}$\cdot$ \EXP{-500}{T} 		& \tablenotemark{18}   \\
88 	& O + HNO $\rightarrow$ OH + NO   						& 5.99\e{-11} 						&  \tablenotemark{18}   \\
89 	& OH + HNO $\rightarrow$ H\2O + NO   						& 8.0\e{-11} $\cdot$ \EXP{-500}{T}		& \tablenotemark{18}   \\
90 	& HNO\2 + OH $\rightarrow$ H\2O + NO\2   					& 1.8\e{-11}$\cdot$ \EXP{-390}{T} 		&  \tablenotemark{1}   \\\hline
91 	& CH$_{4}$ + O $\rightarrow$ CH$_{3}$ + OH   				& 8.32\e{-12}$\cdot$ (T/298)$^{1.56} \cdot$ \EXP{-4300}{T} 		& \tablenotemark{2}   \\
92 	& $^{1}$CH\2 + H\2 $\rightarrow$ CH$_{3}$ + H  				& 1.05\e{-10} 						& \tablenotemark{19}   \\
93 	& $^{1}$CH\2 + CO\2 $\rightarrow$ H\2CO + CO   				& 1\e{-12} 						& \tablenotemark{20}   \\
94 	& $^{3}$CH\2 + O $\rightarrow$ HCO + H   					& 1\e{-11} 							& \tablenotemark{21}   \\
95 	& $^{3}$CH\2 + CO\2 $\rightarrow$ H\2CO + CO   				& 3.9\e{-14} 						& \tablenotemark{3}   \\\hline
96 	& C\2H$_{6}$ + OH $\rightarrow$ C\2H$_{5}$ + H\2O   			& 8.5\e{-13}$\cdot$ (T/298)$^{2.22} \cdot$ \EXP{373}{T} 		&  \tablenotemark{22}   \\
97 	& C\2H$_{6}$ + O $\rightarrow$ C\2H$_{5}$ + OH   				& 8.54\e{-12}$\cdot$ (T/298)$^{1.5} \cdot$ \EXP{-2920}{T} 		&  \tablenotemark{2}   \\
98 	& C\2H$_{6}$ + O($^{1}$D) $\rightarrow$ C\2H$_{5}$ + OH   		& 3.4\e{-10}						& \tablenotemark{23}   \\
99 	& C\2H$_{5}$ + H $\rightarrow$ CH$_{3}$ + CH$_{3}$   			& 1.25\e{-10}				 		& \tablenotemark{24}   \\
100 	& C\2H$_{5}$ + O $\rightarrow$ CH$_{3}$ + HCO + H 			& 1.1\e{-10}						& \tablenotemark{2}   \\\hline
101 	& C\2H$_{5}$ + OH $\rightarrow$ CH$_{3}$ + HCO + H\2 		& As R68 							& \tablenotemark{e}   \\
102 	& C\2H$_{5}$ + HCO $\rightarrow$ C\2H$_{6}$ + CO   			& 3.01\e{-11} 						& \tablenotemark{3} \tablenotemark{f}   \\
103 	& C\2H$_{5}$ + HNO $\rightarrow$ C\2H$_{6}$ + NO   			& 1.66\e{-12} 						& \tablenotemark{25}   \\
104 	& C\2H$_{5}$ + O\2 $\rightarrow$ CH$_{3}$ + HCO + OH 		& \tbdy{1.5\e{-28}}{-3.0}{8.0\e{-12}}{0} 	&  \tablenotemark{16}   \tablenotemark{b} \\
105 	& SO\2 + h$\nu$ $\rightarrow$ SO + O   						& 8.3\e{-5}; 2.1\e{-5}					&  \tablenotemark{c}  \\\hline
106 	& SO + O\2 $\rightarrow$ O + SO\2   						& 1.25\e{-13}$\cdot$ \EXP{-2190}{T} 	& \tablenotemark{1}   \\
107 	& SO + HO\2 $\rightarrow$ SO\2 + OH   						& 2.8\e{-11} 						&  \tablenotemark{1}   \\
108 	& SO + O + M $\rightarrow$ SO\2 + M   						& 6.0\e{-31}$\cdot$ [M] 				& \tablenotemark{27} \tablenotemark{e}  \\
109 	& SO + OH $\rightarrow$ SO\2 + H   						& 8.6\e{-11} 						&  \tablenotemark{16}  \\
110 	& SO\2 + OH + M $\rightarrow$ HSO$_{3}$ + M   				& \tbdy{3\e{-31}}{-3.3}{1.5\e{-12}}{0} 		&  \tablenotemark{16}  \tablenotemark{b} \\\hline
111 	& SO\2 + O + M $\rightarrow$ SO$_{3}$ + M   					& \tbdy{1.8\e{-33}}{-2}{4.2\e{-14}}{-1.8}	& \tablenotemark{1}  \\
112 	& SO$_{3}$ + H\2O + M $\rightarrow$ H\2SO$_{4}$ + M   		& 6.0\e{-15} 						&  \tablenotemark{26}   \\
113 	& HSO$_{3}$ + O\2 $\rightarrow$ HO\2 + SO$_{3}$   			& 1.3\e{-12}$\cdot$\EXP{-330}{T} 		&  \tablenotemark{26}   \\
114 	& HSO$_{3}$ + OH $\rightarrow$ H\2O + SO$_{3}$   			& 1.0\e{-11} 						& \tablenotemark{27} \tablenotemark{e}   \\
115 	& HSO$_{3}$ + H $\rightarrow$ H\2 + SO$_{3}$   				& 1.0\e{-11} 						& \tablenotemark{27} \tablenotemark{e}   \\\hline
116 	& HSO$_{3}$ + O $\rightarrow$ OH + SO$_{3}$   				& 1.0\e{-11} 						& \tablenotemark{27} \tablenotemark{e}   \\
117 	& SO\2 + h$\nu$ $\rightarrow$ SO$_{2}^{1}$     				& 9.4\e{-4}; 2.6\e{-6}					&  \tablenotemark{c}  \\
118 	& SO\2 + h$\nu$ $\rightarrow$ SO$_{2}^{3}$     				& 5.1\e{-7}; 1.9\e{-8}					& \tablenotemark{c}   \\
119 	& SO$_{3}$ + h$\nu$ $\rightarrow$ SO\2 + O   					& 0; 0							& \tablenotemark{c}   \\
120 	& SO$_{2}^{1}$ + M $\rightarrow$ SO$_{2}^{3}$ + M   			& 1.0\e{-12} 						& \tablenotemark{28}   \\\hline
121 	& SO$_{2}^{1}$ + M $\rightarrow$ SO\2 + M   					& 1.0\e{-11} 						& \tablenotemark{28}   \\
122 	& SO$_{2}^{1}$   $\rightarrow$ SO$_{2}^{3}$ + h$\nu$   			& 1.5\e{3} 							& \tablenotemark{28}   \\
123 	& SO$_{2}^{1}$   $\rightarrow$ SO\2 + h$\nu$   				& 2.2\e{4} 							& \tablenotemark{28}   \\
124 	& SO$_{2}^{1}$ + O\2 $\rightarrow$ SO$_{3}$ + O   				& 1.0\e{-16} 						& \tablenotemark{28}   \\
125 	& SO$_{2}^{1}$ + SO\2 $\rightarrow$ SO$_{3}$ + SO   			& 4.0\e{-12} 						& \tablenotemark{28}   \\\hline
126 	& SO$_{2}^{3}$ + M $\rightarrow$ SO\2 + M   					& 1.5\e{-13} 						& \tablenotemark{28}   \\
127 	& SO$_{2}^{3}$   $\rightarrow$ SO\2 + h$\nu$   				& 1.13\e{3} 						& \tablenotemark{28}   \\
128 	& SO$_{2}^{3}$ + SO\2 $\rightarrow$ SO$_{3}$ + SO   			& 7.0\e{-14} 						& \tablenotemark{28}   \\
129 	& SO + NO\2 $\rightarrow$ SO\2 + NO   						& 1.4\e{-11} 						&  \tablenotemark{1}   \\
130 	& SO + O$_{3}$ $\rightarrow$ SO\2 + O\2   					& 3.6\e{12}$\cdot$ \EXP{-1100}{T} 		&  \tablenotemark{1}   \\\hline
131 	& SO\2 + HO\2 $\rightarrow$ SO$_{3}$ + OH   					& 1.0\e{-18} 						& \tablenotemark{29} \tablenotemark{e}   \\
132 	& SO$_{3}$ + SO $\rightarrow$ SO\2 + SO\2   					& 2.0\e{-15} 						& \tablenotemark{30}   \\
133 	& SO + HO\2 $\rightarrow$ HSO + O\2   						& 2.8\e{-11} 						&  \tablenotemark{26}   \\
134 	& SO + HCO $\rightarrow$ HSO + CO   						& As R44 							& \tablenotemark{e}   \\
135 	& H + SO + M $\rightarrow$ HSO + M   						& As R6 							& \tablenotemark{e}   \\\hline
136 	& HSO + NO $\rightarrow$ HNO + SO   						& 1.0\e{-15}						& \tablenotemark{1}   \\
137 	& HSO + OH $\rightarrow$ H\2O + SO   						& As R11 							& \tablenotemark{e}   \\
138 	& HSO + H $\rightarrow$ H\2 + SO   							& As R7 							& \tablenotemark{e}   \\
139 	& HSO + O $\rightarrow$ OH + SO   							& As R13 							& \tablenotemark{e}   \\
140 	& $^{1}$CH\2 + O\2 $\rightarrow$ H\2CO + O   					& 3.0\e{-11} 						& \tablenotemark{8}   \\\hline
141 	& $^{3}$CH\2 + O\2 $\rightarrow$ H\2CO + O   					& 0 								& \tablenotemark{e}   \\
142 	& C\2H$_{6}$ + h$\nu$ $\rightarrow$ CH$_{3}$ + CH$_{3}$   		& 1.2\e{-7}; 4.3\e{-7}					& \tablenotemark{c}   \\
143 	& CH$_{4}$ + h$\nu$ $\rightarrow$ $^{3}$CH\2 + H + H 			& 6.7\e{-7}; 2.3\e{-6}					& \tablenotemark{c}   \\
144 	& CH$_{4}$ + h$\nu$ $\rightarrow$ CH$_{3}$ + H   				& 1.4\e{-6}; 4.8\e{-6}					&  \tablenotemark{c}  \\
145 	& $^{3}$CH\2 + O $\rightarrow$ CO + H + H 					& 8.3\e{-11} 						& \tablenotemark{28}   \\\hline
146 	& $^{3}$CH\2 + H + M $\rightarrow$ CH$_{3}$ + M   			& \tbdyt{3.1\e{-30}$\cdot$ \EXP{457}{T}}{1.5\e{-10}} & \tablenotemark{29} \tablenotemark{d}   \\
147 	& CH$_{3}$ + O\2 + M $\rightarrow$ CH$_{3}$O\2 + M   			& \tbdy{4.0\e{-30}}{-3.6}{1.2\e{-12}}{1.1} 	&  \tablenotemark{1}  \tablenotemark{b} \\
148 	& CH$_{3}$ + H\2CO $\rightarrow$ CH$_{4}$ + HCO   			& 6.8\e{-12}$\cdot$ \EXP{-4450}{T} 		& \tablenotemark{2}   \\
149 	& CH$_{3}$ + OH $\rightarrow$ CO + H\2 + H\2				& 9.1\e{-11}$\cdot$\EXP{-1500}{T}	& \tablenotemark{11} \tablenotemark{f}   \\
150 	& CH$_{3}$O\2 + H $\rightarrow$ CH$_{4}$ + O\2   				& 1.17\e{-11}$\cdot$ (T/298 K)$^{1.02}$$\cdot$ \EXP{-8350}{T} & \tablenotemark{34}   \\\hline
151 	& CH$_{3}$O\2 + H $\rightarrow$ H\2O + H\2CO   				& 1.0\e{-11} 						&  \tablenotemark{3} \tablenotemark{f}  \\
152 	& CH$_{3}$O\2 + O $\rightarrow$ H\2CO + HO\2   				& 1.0\e{-11} 						& \tablenotemark{27} \tablenotemark{e}   \\
153 	& CH$_{3}$CO + H $\rightarrow$ CH$_{4}$ + CO   				& 1.0\e{-10} 						& \tablenotemark{20}   \\
154 	& CH$_{3}$CO + O $\rightarrow$ H\2CO + HCO   				& 5.0\e{-11} 						& \tablenotemark{20}   \\
155 	& CH$_{3}$CO + CH$_{3}$ $\rightarrow$ C\2H$_{6}$ + CO   		& 5.4\e{-11} 						& \tablenotemark{35}   \\\hline
156 	& C\2H$_{5}$ + OH $\rightarrow$ CH$_{3}$ + HCO + H\2 		& As R68 							& \tablenotemark{27} \tablenotemark{e}   \\
157 	& C\2H$_{5}$ + O $\rightarrow$ CH$_{3}$ + HCO + H 			& As R68 							& \tablenotemark{27} \tablenotemark{e}   \\
158 	& C\2H$_{5}$ + H + M $\rightarrow$ C\2H$_{6}$ + M   			& 2.25\e{-10}$\cdot$ (T/298)$^{0.16}$ 	& \tablenotemark{36} \\
159 	& $^{1}$CH\2 + H\2 $\rightarrow$ $^{3}$CH\2 + H\2   			& 1.05\e{-11} 						& \tablenotemark{19}   \\
160 	& HCO + H\2CO $\rightarrow$ CH$_{3}$O + CO   				& 3.8\e{-17} 						& \tablenotemark{37}   \\\hline
161 	& CH$_{3}$O + CO $\rightarrow$ CH$_{3}$ + CO\2   			& 2.6\e{-11}$\cdot$ \EXP{-5940}{T} 		& \tablenotemark{37}   \\
162 	& H\2S + OH $\rightarrow$ H\2O + HS   						& 6.0\e{-12}$\cdot$ \EXP{-75}{T} 		&  \tablenotemark{1}   \\
163 	& H\2S + H $\rightarrow$ H\2 + HS   							& 6.6\e{-11}$\cdot$ \EXP{-1350}{T} 		&  \tablenotemark{38}   \\
164 	& H\2S + O $\rightarrow$ OH + HS   							& 9.2\e{-12}$\cdot$ \EXP{-1800}{T} 		&  \tablenotemark{16}  \\
165 	& HS + O $\rightarrow$ H + SO   							& 1.6\e{-10} 						&  \tablenotemark{16}  \\\hline
166 	& HS + O\2 $\rightarrow$ OH + SO   						& 4.0\e{-19} 						&  \tablenotemark{16}   \\
167 	& HS + HO\2 $\rightarrow$ H\2S + O\2   						& 1.0\e{-11} 						& \tablenotemark{39}   \\
168 	& HS + HS $\rightarrow$ H\2S + S   							& 1.2\e{-11} 						&  \tablenotemark{40}   \\	
169 	& HS + HCO $\rightarrow$ H\2S + CO   						& 5.0\e{-11} 						& \tablenotemark{e}   \\
170 	& HS + H $\rightarrow$ H\2 + S   							& 1.0\e{-11} 						& \tablenotemark{39}   \\\hline
171 	& HS + H\2CO $\rightarrow$ H\2S + HCO   					& 1.7\e{-11}$\cdot$ \EXP{-800}{T} 		&  \tablenotemark{26}  \\
172 	& HS + O$_{3}$ $\rightarrow$ HSO + O\2   					& 9.0\e{-12}$\cdot$ \EXP{-280}{T} 		&  \tablenotemark{41}  \\
173 	& HS + NO\2 $\rightarrow$ HSO + NO   						& 2.9\e{-11}$\cdot$ \EXP{240}{T} 		&  \tablenotemark{41}   \\
174 	& S + O\2 $\rightarrow$ SO + O   							& 2.3\e{-12} 						&  \tablenotemark{41}   \\
175 	& S + OH $\rightarrow$ SO + H   							& 6.6\e{-11} 						&  \tablenotemark{16}   \\\hline
176 	& S + HCO $\rightarrow$ HS + CO   							& 5.0\e{-11} 						& \tablenotemark{e}   \\
177 	& S + HO\2 $\rightarrow$ HS + O\2  							& 1.5\e{-11} 						& \tablenotemark{e}   \\
178 	& S + HO\2 $\rightarrow$ SO + OH   						& 1.5\e{-11} 						& \tablenotemark{e}   \\
179 	& S + O$_{3}$ $\rightarrow$ SO + O\2   						& 1.2\e{-11} 						& \tablenotemark{41}   \\
180 	& S + CO\2 $\rightarrow$ SO + CO   						& 1.2\e{-11} 						& \tablenotemark{30}   \\\hline
181 	& HSO + H $\rightarrow$ HS + OH   							& As R9 							& \tablenotemark{e}   \\
182 	& HSO + HS $\rightarrow$ H\2S + SO   						& 1.0\e{-12} 						& \tablenotemark{e}   \\	
183 	& HSO + S $\rightarrow$ HS + SO   							& 1.0\e{-11} 						& \tablenotemark{e}   \\
184 	& HSO + h$\nu$ $\rightarrow$ HS + O   						& As R50							&  \tablenotemark{c}  \\
185 	& H\2S + h$\nu$ $\rightarrow$ HS + H   						& 1.4\e{-4}; 6.8\e{-7}					&  \tablenotemark{c}  \\\hline
186 	& NH$_{3}$ + h$\nu$ $\rightarrow$ NH\2 + H   					& 5.2\e{-5}; 3.7\e{-7}					&  \tablenotemark{c}  \\
187 	& NH$_{3}$ + OH $\rightarrow$ NH\2 + H\2O   					& 1.60\e{-13}						&  \tablenotemark{41}   \\
188 	& NH$_{3}$ + O($^{1}$D) $\rightarrow$ NH\2 + OH  		 		& 2.5\e{-10} 						&  \tablenotemark{16}   \\
189 	& NH\2 + H + M $\rightarrow$ NH$_{3}$ + M   					& (6.5\e{-30}$\cdot$ [M])/ (1 + 3\e{-20}$\cdot$ [M]) & \tablenotemark{42}   \\
190 	& NH\2 + NO $\rightarrow$ N\2 + H\2O   						& 2.07\e{-11}$\cdot$ (T/298)$^{-1.61} \cdot$ \EXP{-150}{T} 		&  \tablenotemark{43}   \\\hline
191 	& NH\2 + NH\2 + M $\rightarrow$ N\2H$_{4}$ + M   				& 1.96\e{-29}$\cdot$ (T/298)$^{-3.9}$	& \tablenotemark{44}   \\
192 	& NH\2 + O $\rightarrow$ NH + OH   						& 1.16\e{-11}						& \tablenotemark{45}   \\
193 	& NH\2 + O $\rightarrow$ HNO + H   						& 7.47\e{-11} 						& \tablenotemark{45}   \\
194 	& NH + NO $\rightarrow$ N\2 + O + H 						& 4.9\e{-11} 						&  \tablenotemark{16} \tablenotemark{f}   \\
195 	& NH + O $\rightarrow$ N + OH   							& 1.0\e{-11} 						& \tablenotemark{45}   \\\hline
196 	& N\2H$_{4}$ + h$\nu$ $\rightarrow$ N\2H$_{3}$ + H   			& 9.3\e{-5}; 6.5\e{-7}					& \tablenotemark{c}   \\
197 	& N\2H$_{4}$ + H $\rightarrow$ N\2H$_{3}$ + H\2   				& 9.9\e{-12}$\cdot$ \EXP{-1200}{T} 		& \tablenotemark{46}   \\
198 	& N\2H$_{3}$ + H $\rightarrow$ NH\2 + NH\2   					& 2.7\e{-12} 						& \tablenotemark{47}  \\
199 	& N\2H$_{3}$ + N\2H$_{3}$ $\rightarrow$ N\2H$_{4}$ + N\2 + H\2 	& 6.0\e{-11} 						& \tablenotemark{48}   \\
200 	& NH + H + M $\rightarrow$ NH\2 + M   						& As R190 						& \tablenotemark{e}   \\\hline
201 	& NH + h$\nu$ $\rightarrow$ N + H   						& As R186 						& \tablenotemark{e}   \\
202 	& NH\2 + h$\nu$ $\rightarrow$ NH + H   						& As R186 						& \tablenotemark{e}    \\
203 	& NH\2 + h$\nu$ $\rightarrow$ $^{X}$NH\2     					& 3.8\e{-3} 						& \tablenotemark{49}   \\
204 	& $^{X}$NH\2   $\rightarrow$ NH\2 + h$\nu$   					& 1.2\e{5} 							& \tablenotemark{50}  \\
205 	& $^{X}$NH\2 + M $\rightarrow$ NH\2 + M   					& 3.0\e{-11} 						& \tablenotemark{e}  \\\hline
206 	& $^{X}$NH\2 + H\2 $\rightarrow$ NH$_{3}$ + H   				& 3.0\e{-11} 						& \tablenotemark{e}   \\
207 	& NH\2 + HCO $\rightarrow$ NH$_{3}$ + CO   					& 1.0\e{-11} 						& \tablenotemark{e}   \\
208 	& NH + HCO $\rightarrow$ NH\2 + CO   						& 1.0\e{-11} 						& \tablenotemark{e}   \\
209 	& NO + h$\nu$ $\rightarrow$ N + O   						& 1.9\e{-6}; 1.9\e{-6}					&  \tablenotemark{c}  \\
210 	& N + N + M $\rightarrow$ N\2 + M   							& 1.0\e{-10} 						& \tablenotemark{e}   \\\hline
211 	& N + O\2 $\rightarrow$ NO + O   							& 1.5\e{-11}$\cdot$ \EXP{-3600}{T} 		&  \tablenotemark{16}   \\
212 	& N + O$_{3}$ $\rightarrow$ NO + O\2   						& 1.0\e{-10} 						& \tablenotemark{e}   \\
213 	& N + OH $\rightarrow$ NO + H   							& 4.7\e{-11} 						& \tablenotemark{15}   \\	
214 	& N + NO $\rightarrow$ N\2 + O   							& 2.19\e{-11}$\cdot$\EXP{160}{T}		&  \tablenotemark{51}    
\enddata

\end{deluxetable}
\clearpage

Notes--- References: \tablenotemark{1} \citet{sander2011}; \tablenotemark{2} \citet{baulch1992} [NIST]; \tablenotemark{3} \citet{tsang1986} [NIST]; \tablenotemark{4} \citet{friedrichs2002} [NIST]; \tablenotemark{5} \citet{pinto1980}; \tablenotemark{6} \citet{oehlers2000} [NIST]; \tablenotemark{7} \citet{yung1999}; \tablenotemark{8} \citet{ashfold1981}; \tablenotemark{9} \citet{tian2014}; \tablenotemark{10} \citet{prasad1980}; \tablenotemark{11} \citet{humpfer1995} [NIST]; \tablenotemark{12} \citet{wagner1988}; \tablenotemark{13} \citet{brouard1989}; \tablenotemark{14} \citet{choi2005} [NIST]; \tablenotemark{15} \citet{baulch1994} [NIST]; \tablenotemark{16} \citet{demore1997} [NIST]; \tablenotemark{17} \citet{veyret1981} [NIST]; \tablenotemark{18} \citet{tsang1991} [NIST];   \tablenotemark{19} \citet{gannon2008} [NIST]; \tablenotemark{20} \citet{zahnle1986}; \tablenotemark{21} \citet{huebner1980}; \tablenotemark{22} \citet{krasnoperov2004} [NIST]; \tablenotemark{23} \citet{dillon2007} [NIST]; \tablenotemark{24} \citet{sillesen1993} [NIST]; \tablenotemark{25} \citet{laidler1961} [NIST];  \tablenotemark{26} \citet{demore1992}; \tablenotemark{27} \citet{kasting1990};  \tablenotemark{28} \citet{turco1982}; \tablenotemark{29} \citet{graham1979}; \tablenotemark{30} \citet{yung1982}; \tablenotemark{32} \citet{homan1981}; \tablenotemark{33} \citet{gladstone1983};  \tablenotemark{34} \citet{bogdanchikov2004} [NIST]; \tablenotemark{35} \citet{adachi1986}; \tablenotemark{36} \citet{harding2005} [NIST]; \tablenotemark{37} \citet{wen1989}; \tablenotemark{38} \citet{peng1999} [NIST]; \tablenotemark{39} \citet{stachnik1987} [NIST]; \tablenotemark{40} \citet{baulch1976}; \tablenotemark{41} \citet{atkinson2004} [NIST]; \tablenotemark{42} \citet{gordon1971}; \tablenotemark{42} \citet{park1999} [NIST]; \tablenotemark{44} \citet{fagerstrom1995} [NIST]; \tablenotemark{45} \citet{cohen1991} [NIST];  \tablenotemark{46} \citet{stief1976}; \tablenotemark{47} \citet{gehring1969} [NIST]; \tablenotemark{48} \citet{kuhn1979}; \tablenotemark{49} \citet{kasting1982}; \tablenotemark{50} \citet{lenzi1972};  \tablenotemark{51} \citet{wennberg1994} [NIST]. 

$^{a}$ These reactions are particularly sensitive to the chosen rate constant.

$^{b}$ These reaction rates take the form: k(M,T) =  $k_{0}(T)[M]/[1+k_{0}(T)[m]/k_{\infty}(T)]\cdot$ 0.6\^{}$[1+[log_{10}[k_{0}(T)[M]/k_{\infty}(T)]]^{2}]^{-1}$, where $k_{0}(T)$ has units of cm$^{6}$ molecules$^{-2}$ s$^{-1}$ and k$_{\infty}(T)$ has units of cm$^{3}$ molecules$^{-2}$ s$^{-1}$.

$^{c}$ The photolysis rates presented here are taken from the uppermost layer in the model ($\sim$100 km) for the Sun (first value) and GJ 876 (second value). Caution: the rates in the upper atmosphere are not good indicators for rates in the lower atmosphere.

$^{d}$ These reaction rates take the form: k(M,T) = $[k_{0}(T)k_{\infty}(T)[M]]/[k_{0}(T)[M] + k_{\infty}(T)]$, where k$_{0}$ has units of cm$^{6}$ molecules$^{-2}$ s$^{-1}$ and k$_{\infty}(T)$ has units of cm$^{3}$ molecules$^{-2}$ s$^{-1}$. If FC is given, use instead: log$_{10}$(k) = $log_{10}[k_{0}(T)/[1 + k_{0}(T)/k_{\infty}(T)]] + (log_{10}(FC)/[1 + log_{10}(k_{0}(T)/k_{\infty}(T))^{2}]$, following \citet{wagner1988}.

$^{e}$ Estimated rate constant or branching ratio. 

$^{f}$ Modified products.

\bibliographystyle{apj}
\bibliography{o2refs}

\clearpage

\end{document}